\documentclass[preprint,12pt]{elsarticle}

\PassOptionsToPackage{numbers, square}{natbib}

\usepackage{amssymb}
\usepackage{amsmath}

\usepackage{array}
\usepackage{multirow}
\usepackage{booktabs}
\usepackage{tabularx} 
\usepackage{tcolorbox}
\usepackage{graphicx}
\tcbuselibrary{breakable}
\usepackage{subcaption} 
\usepackage{hyperref}
\usepackage{makecell} 
\usepackage{stackengine}
\usepackage{booktabs}
\usepackage{tabularx}
\usepackage{tcolorbox}
\usepackage{graphicx}
\usepackage{caption} 
\usepackage{lipsum} 
\usepackage{float}
\usepackage{longtable}
\usepackage{indentfirst}
\journal{Nuclear Physics B}

\begin{document}

\begin{frontmatter}



\title{MaintAGT:Sim2Real-Guided Multimodal Large Model for Intelligent Maintenance with Chain-of-Thought Reasoning} 


\author[label1]{Hongliang He}
\author[label1]{Jinfeng Huang}
\author[label1]{Qi Li}
\author[label1]{Xu Wang}
\author[label1]{Feibin Zhang\corref{cor1}}
\author[label2]{Kangding Yang}
\author[label2]{Li Meng}
\author[label1]{Fulei Chu\corref{cor2}}

\cortext[cor1]{Corresponding author: Feibin Zhang. Email: zfbin2008@163.com}
\cortext[cor2]{Corresponding author: Fulei Chu. Email: chufl@mail.tsinghua.edu.cn}
\cortext[cor0]{author: Hongliang He. Email: hhl127@mail.tsinghua.edu.cn}

\affiliation[label1]{organization={The State Key Laboratory of Tribology, Department of Mechanical Engineering},
            addressline={ Tsinghua University}, 
            city={Beijing},
            postcode={100084}, 
            country={China}}

\affiliation[label2]{organization={FreqX Intelligence Technology Co.},
            addressline={Ltd.}, 
            city={Changzhou},
            postcode={213162}, 
            country={China}}

\begin{abstract}

In recent years, large language models have made significant advancements in the field of natural language processing, yet there are still inadequacies in specific domain knowledge and applications. This paper Proposes MaintAGT, a professional large model for intelligent operations and maintenance, aimed at addressing this issue. The system comprises three key components: a signal-to-text model, a pure text model, and a multimodal model. Firstly, the signal-to-text model was designed to convert raw signal data into textual descriptions, bridging the gap between signal data and text-based analysis. Secondly, the pure text model was fine-tuned using the GLM4 model with specialized knowledge to enhance its understanding of domain-specific texts. Finally, these two models were integrated to develop a comprehensive multimodal model that effectively processes and analyzes both signal and textual data.The dataset used for training and evaluation was sourced from academic papers, textbooks, international standards, and vibration analyst training materials, undergoing meticulous preprocessing to ensure high-quality data. As a result, the model has demonstrated outstanding performance across multiple intelligent operations and maintenance tasks, providing a low-cost, high-quality method for constructing large-scale monitoring signal-text description-fault pattern datasets. Experimental results indicate that the model holds significant advantages in condition monitoring, signal processing, and fault diagnosis.In the constructed general test set, MaintAGT achieved an accuracy of 70\%, surpassing all existing general large language models and reaching the level of an ISO Level III human vibration analyst.This advancement signifies a crucial step forward from traditional maintenance practices toward intelligent and AI-driven maintenance solutions.
\end{abstract}

\begin{keyword}



Intelligent Maintenance
\sep Multimodal Models 
\sep Large Language Models
\sep Fault Diagnosis
\sep Condition Monitoring

\end{keyword}

\end{frontmatter}



\section{Introduction}
\label{sec1}
In recent years, the manufacturing industry has experienced a significant transformation toward automation, digitization, and intelligence, leading to the widespread adoption of intelligent manufacturing equipment \cite{equipment_1,equipment_2}. This shift is particularly evident in roles involving high risk, intensity, or repetition, where `machine replacement' has become increasingly prominent.Equipment failure is inevitable, particularly in prolonged and repetitive work environments where intelligent manufacturing equipment is prone to performance degradation and damage \cite{damage_3,damage_4,damage_5,damage_6}. This highlights the urgent need for an intelligent operations and maintenance (O\&M) system capable of proactively monitoring and predicting equipment health and potential faults. Such a system would enable timely maintenance decisions, reduce unexpected downtime, lower production costs, and enhance production safety \cite{safety_7,safety_8}.

Traditional fault diagnosis methods often rely heavily on expert knowledge, making them time-consuming and prone to misdiagnosis \cite{big_data_10}. In contrast, intelligent fault diagnosis methods based on big data analytics and machine learning can directly infer fault types from operational data. Numerous studies have explored the potential of machine learning in intelligent fault diagnosis. Zhang et al. \cite{Zhang_11} Proposed a semi-supervised classifier based on low-rank tensor learning and graph embedding for intelligent diagnostics. While these machine learning-based models have shown promise, their performance largely depends on high-quality, labeled datasets. Creating extensive and accurately labeled datasets is challenging due to data heterogeneity and the significant effort required for labeling, which can limit the models' generalization across different factories \cite{domain_adversarial_training_12,deep_adversarial_domain_13}.

The advent of large language models (LLMs) represents a significant advancement in machine learning and deep learning algorithms \cite{llms1,llms2}, offering timely opportunities to address these challenges. LLMs like GPT-4 , ChatGPT \cite{chatgpt4,chatgpt5}, and open-source models such as LLaMA \cite{llama7,llama8,llama9} have demonstrated exceptional capabilities in natural language processing, enabling interactive and contextually coherent conversations. By integrating text, images, and other data formats, LLMs facilitate cross-domain applications and enhance AI's problem-solving abilities. Domain-specific LLMs have also emerged to meet the unique needs of different sectors, such as BloombergGPT \cite{bloomberg_gpt} , in finance, LawGPT \cite{law} in legal services, and BioGPT \cite{biogpt} in biomedical research. These advancements highlight LLMs' remarkable comprehension and reasoning abilities across various domains, demonstrating their potential versatility and adaptability.

Recognizing the potential of LLMs in fault diagnosis and addressing existing research gaps, several studies have attempted to apply LLMs in intelligent fault diagnosis. For instance, Sun et al.\cite{sun} developed an LLM-driven Digital Twin architecture to enhance fault diagnosis by leveraging multi-modal data and addressing gaps in adaptability and interpretability. Xia et al. \cite{xia} proposed an error-assisted fine-tuning framework for adapting large language models (LLMs) to the manufacturing domain, enabling accurate response generation and reliable code for improved manufacturing applications. However, these approaches often face limitations, such as reliance on annotated industry datasets and inability to process multimodal data effectively. Zheng et al. \cite{zheng} proposed an LLM-based fault diagnosis model where normalized high-speed rail experimental data served as prompt samples for model fine-tuning, enabling sample classification. Nevertheless, their data processing methodology closely resembled traditional intelligent fault diagnosis, heavily depending on annotated datasets and lacking advanced reasoning mechanisms like chain-of-thought reasoning \cite{CoT_1,CoT_2,CoT_3,CoT_4}. Lowenmark et al. \cite{lowenmark} built a supervised model annotated with industrial signals and technical language, extracting features from unlabeled sensor data and mapping them into annotation embeddings for model training. Yet, these models often fall short in specialized domains, leading to less accurate diagnoses and maintenance recommendations, potentially jeopardizing safety.

In the field of intelligent diagnostics, constructing a professional large model presents numerous challenges. High-quality monitoring signal data are difficult to obtain, particularly fault data, which are much scarcer compared to healthy data, leading to significant data imbalance \cite{imbalance_1,imbalance_2}. Labeling monitoring signal data is challenging due to the specialized expertise required, making it difficult to label data cost-effectively, in large quantities, and with high accuracy and efficiency \cite{efficiency_1,efficiency_2}. Moreover, engineering signal monitoring data typically contain significant noise, and the monitoring signals often consist of multi-source, multi-feature component signals that are linearly or nonlinearly superimposed, making it challenging even for professionals to accurately and quickly label the raw signals.

To fundamentally address these issues, this paper takes an alternative approach by circumventing the direct construction of monitoring signal-fault labels, guided by the technical insights of human experts in intelligent O\&M. We first propose a signal-to-text (Sig2Txt) model driven by simulation-to-reality (Sim2Real) techniques to establish an innovative, low-cost, high-quality, large-scale data construction pipeline from monitoring signals to mathematical features to textual descriptions. We then Proposes a specialized large language model for intelligent O\&M by fine-tuning GLM4\cite{glm4} on a substantial collection of textual materials in this field to enhance its domain-specific capabilities. By integrating the Sig2Txt model with the specialized text-based large language model and incorporating a chain-of-thought reasoning approach, we propose a novel multimodal large model for intelligent O\&M, named MaintAGT.

Compared to monitoring signals and other data types, text offers distinct advantages: (1) Nearly all human knowledge and information can be accurately and comprehensively expressed through text, yielding large-scale textual data in both general and specialized fields; (2) Text sources such as professional books, academic papers, industry reports, national standards, and code provide high-quality, large datasets without the compounded feature issues of signal data; (3) Text data can be easily organized and used to generate large-scale, high-quality labeled datasets comparable to those like ImageNet [36]. However, LLMs trained exclusively on text data cannot directly interpret or analyze raw monitoring data. By integrating the Sig2Txt model, we bridge this gap, enabling the model to process signal data by converting it into textual descriptions, thus enhancing its ability to address tasks in intelligent operations and maintenance.

Finally, we evaluate the performance of this multimodal model through simulations and practical tests, including the ISO Level III vibration analyst questions related to fault diagnosis. Our approach enhances its ability to address tasks in intelligent O\&M, achieving human-level expertise validated by testing on multiple fault categories using accuracy as the performance metric. Its capability with multimodal data is further assessed through real-world case studies.

The main innovations of MaintAGT are as follows:

1) Sim2Real Driven Signal-to-Text Model (Sig2Txt): A groundbreaking model that converts complex signal data into structured textual descriptions enriched with mathematical and physical features. This innovation enables more effective interpretation and analysis of maintenance data, laying the foundation for a robust multimodal architecture.

2) High-Quality Specialized Text Dataset and Professional Text Large Model: The creation of a high-quality specialized text dataset for intelligent operations and maintenance, and the development of a professional text large model based on instruction-tuning techniques applied to a general pre-trained language model. This ensures the model possesses deep domain-specific knowledge.

3) Chain-of-Thought Paradigm Multimodal Large Model: By organically integrating Sig2Txt with the professional text large model and fusing the chain-of-thought architecture, the multimodal model gains enhanced reasoning capabilities. This enables automatic diagnosis and recognition of monitoring signal data, expanding the application scope of large language models in intelligent operations and maintenance.

4) Efficient Training Strategy: Utilizing multi-task learning and optimized training configurations to enhance the model's performance and training efficiency. This includes exploring new research directions such as task fine-tuning and human feedback mechanisms to further improve accuracy and practicality.

\section{signal to text}
\subsection{Textual Description of Monitoring Signals}
In the technical workflow for mechanical equipment maintenance carried out by human experts, the primary task is the meticulous analysis of monitoring signals. In other words, using theoretical knowledge and practical experience, along with mathematical statistics, signal processing, and other methods, experts derive mathematical signal features from monitoring data. These features are crucial for status assessment, fault diagnosis, fault prediction, or maintenance decision-making and are described in structured or unstructured text formats. The Sig2Txt model proposed in this paper essentially serves as a converter between mathematical signal features and their textual descriptions. There are two main technical approaches for implementing Sig2Txt: one based on prior algorithmic knowledge and the other leveraging pre-trained large models.

\begin{figure}[H]
    \centering
    \includegraphics[width=1\linewidth]{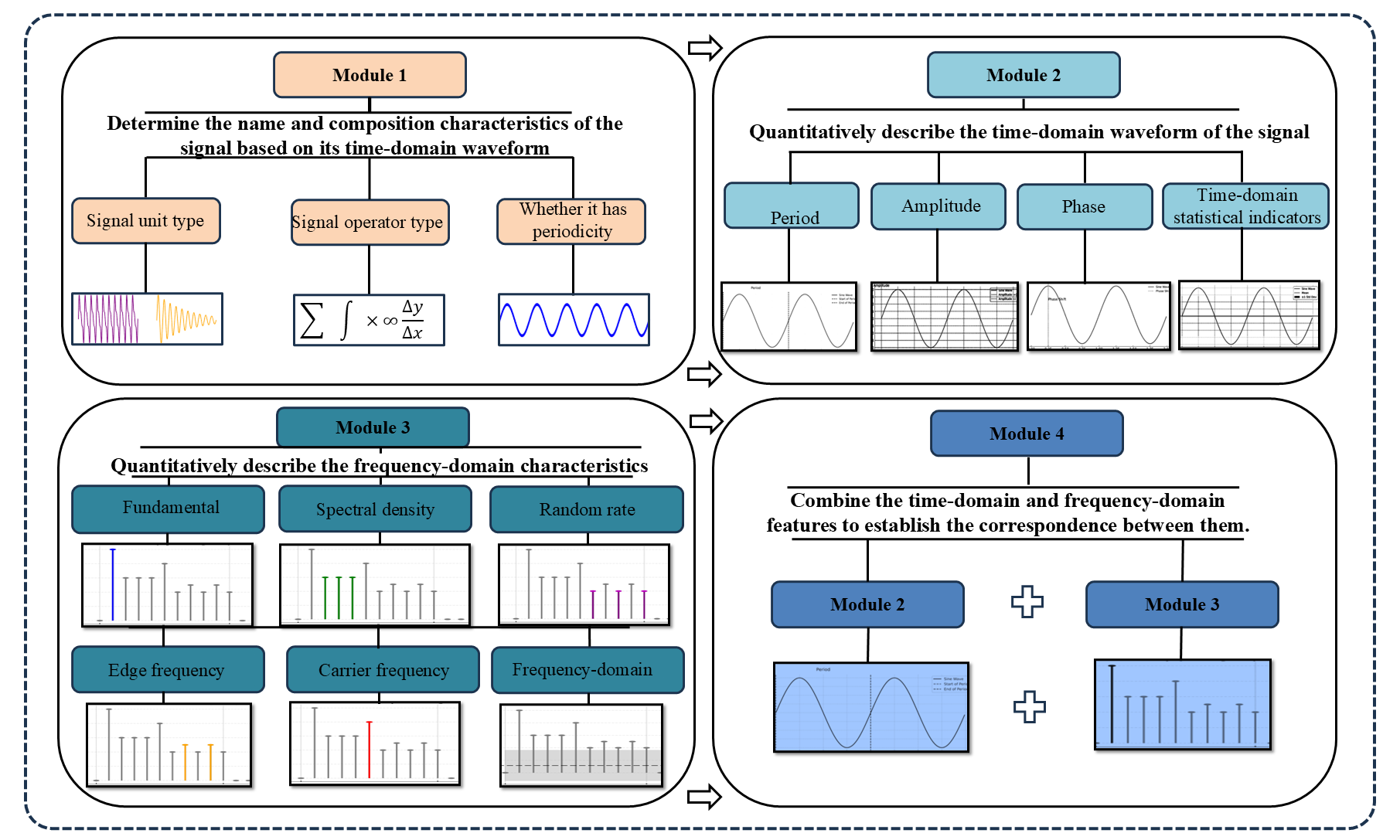}
    \caption{Text description of steps}
    \label{fig:enter-label}
\end{figure}

This chapter illustrates the principles and construction process of the proposed Sig2Txt model using vibration signal analysis as an example. In the field of vibration analysis for machinery maintenance, the primary mathematical signal features to extract include time-domain and frequency-domain characteristics. Table 1 provides a detailed list of typical features for vibration signal analysis. It should be noted that the mathematical features listed in Table 1 do not encompass all the features required for vibration analysis. However, they serve as typical examples, and additional features may be incorporated as needed in real-world applications. Table 1 thus represents the structured textual description of the signal's mathematical features.

\begin{table}[H]
\caption{Extracted features from signals in time and frequency domains}
\label{combined_features}
\begin{tabularx}{\textwidth}{p{3cm} p{3.5cm} X}
\toprule
\textbf{Domain} & \textbf{Type of features} & \textbf{Name of features} \\ \midrule

\multirow{17}{*}{Time domain}
& \multirow{14}{*}{Statistical indicators}
& Root mean square \\
& & Mean \\
& & Kurtosis \\
& & Linear kurtosis \\
& & Margin \\
& & Minimum value \\
& & Maximum value \\
& & Peak to peak value \\
& & Skewness \\
& & Root square amplitude \\
& & Absolute mean \\
& & Variance \\
& & Waveform indicators \\
& & Peak \\ \cmidrule(lr){2-3}
& \multirow{3}{*}{Wave features}
& Fundamental frequency period \\
& & Amplitude modulation period \\
& & Periodic shock \\ \midrule

\multirow{15}{*}{Frequency domain}
& \multirow{9}{*}{Statistical indicators}
& Root mean square \\
& & Kurtosis \\
& & Linear kurtosis \\
& & Center of gravity frequency \\
& & Standard deviation \\
& & Mean value \\
& & Frequency variance \\
& & Frequency standard deviation \\
& & Frequency domain energy \\ \cmidrule(lr){2-3}
& \multirow{6}{*}{Spectrum features}
& 1-N1 octave amplitude of fundamental frequency \\
& & The frequency value with the maximum amplitude of the front N2 \\
& & Center frequency and sideband spacing of amplitude modulation \\
& & Fractional frequency amplitude of fundamental frequency \\
& & 1-N3 harmonic amplitude of non-fundamental frequency \\
& & The maximum center frequency values of the first four amplitudes \\ \bottomrule

\end{tabularx}
\end{table}

\subsection{Signal Denoising Algorithm}
A key driver in transforming vibration signals is to make them interpretable by large language models (LLMs), enabling diverse and customizable downstream tasks. Currently, LLMs function as optimal intelligent assistants, capable of imitating human cognition. They demonstrate reliable single-step inference abilities and show potential for complex multi-step reasoning under effective prompt engineering strategies. This proficiency stems from their training on extensive datasets encompassing both general and specialized knowledge. Moreover, the Transformer architecture enhances their capability for compositional generalization and symbolic reasoning. Through self-attention mechanisms, LLMs adeptly link sequential thoughts, track logical processes, and reach coherent conclusions.

In practical applications, physical equipment often exhibits complex data features across multiple dimensions, with fault data displaying varied behaviors and characteristics across different time scales, along with noise interference[37,38]. A signal denoising algorithm capable of identifying useful features within complex signal data and removing noise interference may yield more accurate diagnostic results. This underscores the need for a signal denoising module in the 'signal-to-text' conversion process, designed to extract and analyze fundamental characteristics of equipment faults while mitigating external noise, ensuring accurate and efficient transformation from raw signals to coherent textual descriptions. Generally, it is challenging for LLMs to directly analyze and learn from noisy vibration signals. A feasible and effective approach is to employ the proposed flexible tensor singular spectrum decomposition method.

\textbf{Definition 1} [Flexible Tensor Product]Let \(\mathcal{A} \in \mathbb{R}^{I_1 \times I_2 \times I_3 \times \cdots \times I_m}\) be an \(I_1 \times I_2 \times I_3 \times \cdots \times I_m\) tensor and \(\mathcal{B} \in \mathbb{R}^{J_1 \times J_2 \times J_3 \times \cdots \times J_n}\) be a \(J_1 \times J_2 \times J_3 \times \cdots \times J_n\) tensor. The flexible tensor product of tensors \(\mathcal{A} \in \mathbb{R}^{I_1 \times I_2 \times I_3 \times \cdots \times I_m}\) and \(\mathcal{B} \in \mathbb{R}^{J_1 \times J_2 \times J_3 \times \cdots \times J_n}\) is given by
\begin{equation}
(\mathcal{A} *_{q \lvert \alpha, \beta \rangle} \mathcal{B})_{k_1,k_2,\cdots,k_{m+n-2q}} = \sum_{s_1=1}^{S_1} \cdots \sum_{s_o=1}^{S_o} \mathcal{A}_{k_1, \overset{\bigstar}{\alpha}} \, \mathcal{B}_{\overset{\bigstar}{\beta}, k_{m+n-2q}}
\end{equation}
where
\begin{equation}
\begin{aligned}
&\begin{cases}
\bigstar_{\alpha} \triangleq s_1 \cdots s_q, k_2\cdots k_{m-q};S_r = I_{r+1}, \, r = 1, 2 \cdots q \text{ when } \alpha = 1\\
\bigstar_{\alpha} \triangleq k_2 \cdots k_{m-q}, s_q \cdots s_1;S_r = I_{m-r+1}, \, r = 1, 2 \cdots q \text{ when } \alpha = 2
\end{cases} \\
&\begin{cases}
\bigstar_{\beta} \triangleq s_1 \cdots s_q, k_{m-q+1} \cdots k_{m+n-2q-1};S_r = J_r, \, r = 1, 2 \cdots q \text{ when } \beta = 1\\
\bigstar_{\beta} \triangleq k_{m-q+1} \cdots k_{m+n-2q-1}, s_q \cdots s_1;S_r = J_{n-r}, \, r = 1, 2 \cdots q \text{ when } \beta = 2
\end{cases}
\end{aligned}
\end{equation}
where \( \alpha \) and \( \beta \), in the notation \(\lvert \alpha, \beta \rangle\), respectively present the order sequence of the tensor \( \mathcal{A} \) and the tensor \( \mathcal{B} \) involved in the flexible tensor product operation, where \( \alpha, \beta \in \{1, 2\} \). Here, `1' denotes the positive sequence, in which the tensor's order follows the forward indexing, while `2'
 denotes the reverse sequence, indicating adherence to the backward indexing. The positive integer parameter \( q \) represents the order of the multiplier in tensor multiplication, with its permissible range \( q \in \{1, 2, \dots, \min(m, n) - 1\} \), where \( m \) and \( n \) are the orders of tensors \( \mathcal{A} \) and \( \mathcal{B} \) respectively.This flexibility in tensor multiplication allows for a broader range of computational possibilities and adapts to various application scenarios.

Following the flexible tensor product principles, the flexible tensor singular value decompositions (SVD) can be derived, which are

\begin{equation}
    \mathcal{A} = \mathcal{U} *_{1 \lvert 1, 1 \rangle} \mathcal{S} *_{1 \lvert 2, 1 \rangle} \mathcal{V}^T
\end{equation}

\begin{equation}
    \mathcal{A} = \mathcal{U} *_{1 \lvert 2, 1 \rangle} \mathcal{S} *_{2 \lvert 1, 1 \rangle} \mathcal{V}^T
\end{equation}

\textbf{Definition 2.1} [The First-kind flexible tensor SVD (1K-FTSVD)]Let \( \mathcal{X} \in \mathbb{R}^{I_1 \times I_2 \times I_3} \) be an \( I_1 \times I_2 \times I_3 \) third-order tensor, the first-kind flexible tensor singular value decomposition is a way to factor \( \mathcal{X} \) as

\begin{equation}
    \mathcal{A} = \mathcal{U} *_{1 \lvert 1, 1 \rangle} \mathcal{S} *_{1 \lvert 2, 1 \rangle} \mathcal{V}^T
\end{equation}
where \( \mathcal{U} \in \mathbb{R}^{I_1 \times I_1} \) and \( \mathcal{V} \in \mathbb{R}^{I_3 \times I_3} \) are orthogonal singular tensors such that \( \mathcal{U}^T \times_{1, \lvert 2, 1 \rangle} \mathcal{U} \equiv I_{I_1} \) and \( \mathcal{V}^T *_{1, \lvert 2, 1 \rangle} \mathcal{V} \equiv I_{I_3} \). The \( \mathcal{S} \in \mathbb{R}^{I_1 \times I_2 \times I_3} \) is a core tensor.

\textbf{Definition 2.2} [The second-kind flexible tensor SVD (2K-FTSVD)]Let \( \mathcal{X} \) be an \( I_1 \times I_2 \times I_3 \) tensor. Then, \( \mathcal{X} \) can be decomposed as

\begin{equation}
    \mathcal{A} = \mathcal{U} *_{1 \lvert 2, 1 \rangle} \mathcal{S} *_{2 \lvert 1, 1 \rangle} \mathcal{V}^T
\end{equation}
where \( \mathcal{U} \) is an \( I_1 \times I_2 \) tensor, \( \mathcal{V} \) is an \( I_2 \times I_3 \times I_2 \times I_3 \) tensor, and \( \mathcal{S} \) is an \( I_1 \times I_2 \times I_3 f\)-diagonal tensor such that each frontal slice is a diagonal matrix. The decomposition (\textbf{Definition 2.2}) is referred to as the second-kind flexible tensor SVD (2K-FTSVD). Analogous to the SVD of matrices, 2K-FTSVD possesses many desirable properties as well:

(1) Pseudo-diagonal: \( \text{mat}_{o,<p,1>}(\mathcal{S}) = \text{diag}(\sigma_1, \sigma_2, \dots, \sigma_i, \dots, \sigma_l) \);

(2) Ordering: \( \sigma_1 \geq \sigma_2 \geq \sigma_3 \geq \dots \geq 0 \);
where the diagonal entries \( \sigma_i \) of the core tensor \( \mathcal{S} \) are called the singular values.

According to the definition 2 and the equation, \textbf{Algorithm 1} gives the Matlab pseudo code for the flexible tensor singular value decompositions (SVD) of a third-order tensor. 

\begin{table}[H]
\centering
\caption{Flexible Tensor Singular Value Decomposition}
\begin{tabular}{l}
\hline
\textbf{Algorithm 1.} The flexible tensor SVD \\
\hline
\textbf{Input:} \( I_1 \times I_2 \times I_3 \) Tensor \( \mathcal{A} \in \mathbb{R}^{I_1 \times I_2 \times I_3} \). \\
\textbf{Output:} \( \mathcal{U} \in \mathbb{R}^{I_1 \times I_1} \), \( \mathcal{S} \in \mathbb{R}^{I_1 \times I_2 \times I_3} \), \( \mathcal{V} \in \mathbb{R}^{I_3 \times I_3} \) or \( \mathcal{V} \in \mathbb{R}^{I_3 \times I_2 \times I_2 \times I_3} \). \\
The first kind: \\
\quad \([ \mathcal{U}, \sim ] = \text{eig} \left( \mathcal{A} *_{2 \lvert 2, 1 \rangle} \mathcal{A}^T \right);\) \\
\quad \([ \mathcal{V}, \sim ] = \text{eig} \left( \mathcal{A}^T *_{2 \lvert 2, 1 \rangle} \mathcal{A} \right);\) \\
\quad \( \mathcal{S} = \mathcal{U}^T *_{1 | (1,1)} \mathcal{A} *_{1 \lvert 1, 1 \rangle} \mathcal{V}; \) \\
The second kind: \\
\quad \( R = \text{unfold}(\mathcal{A}) \) \\
\quad \([ \mathcal{U}, \mathcal{S}, \mathcal{V}^T ] = \text{SVD}[R] \) \\
\quad \( \mathcal{U} = \text{fold}(\mathcal{U}); \) \\
\quad \( \mathcal{S} = \text{fold}(\mathcal{S}); \) \\
\quad \( \mathcal{V}^T = \text{fold}(\mathcal{V}^T); \) \\
\hline
\end{tabular}
\end{table}

After constructing the signal to be analyzed into a ISO Level III Hankel matrix and performing singular value decomposition (SVD) on it, the matrices \( \mathcal{U} \), \( \mathcal{S} \), and \( \mathcal{V} \) are obtained. Then, in matrix \( \mathcal{U} \), we find the components within the frequency band \( [f_{\max} - \Delta f, f_{\max} + \Delta f] \), which corresponds to all the feature components with significant dominant frequencies within this range. We select the feature components that contribute the most energy to the peak spectrum and denote them as \( I_j \) (\( I_j = \{i_1, \cdots, i_k\} \)), where \( \Delta f \) represents the half-width of the peak in the time-series power spectral density. Subsequently, based on the set \( I_j \), we extract the required feature quantities and singular values from the matrices \( \mathcal{U} \), \( \mathcal{S} \), and \( \mathcal{V} \). Next, using the weighted average method and cross-angle averaging, we reconstruct the weighted sum of the singular values back into the signal. The above steps constitute the method for decomposing the singular spectrum of the flexible Hankel matrix.

For the signals to be analyzed, a flexible tensor singular spectrum decomposition method is used to capture their key features and eliminate noise effects, enabling clearer feature extraction in textual descriptions. In the field of vibration analysis for the operation and maintenance of mechanical equipment, the primary mathematical features to be extracted from signals include the time-domain and frequency-domain characteristics, as shown in Table 2.

\subsection{Sig2Txt Based on Prior Algorithm Knowledge}
Large language models primarily process text data in natural language format. While many vibration characteristics can be easily translated into natural language, describing the shape features of signals often involves manual work, resulting in inconsistencies and a lack of standardization. To address this issue, this paper Proposes Sig2Txt (Signal-to-Text Conversion) based on prior algorithm knowledge. First, the signal denoising algorithm—flexible tensor singular spectrum decomposition, discussed in the previous section—is applied to complex signal data to obtain denoised time-domain and frequency-domain results. Next, feature extraction is performed on these time-domain and frequency-domain results, focusing on the characteristics outlined in Table 1. Finally, standardized descriptions of shape features are autonomously generated through text generation rules.

\begin{figure}[H]
    \centering
    \includegraphics[width=1\linewidth]{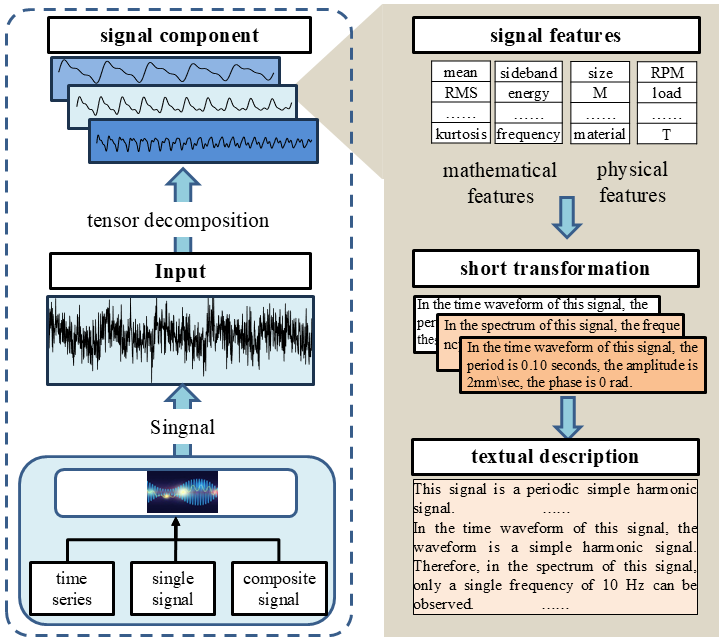}
    \caption{The flowchart of Sig2Txt based on existing algorithmic knowledge}
    \label{fig:enter-label}
\end{figure}

\begin{equation}
    f_{Temp}^{Sig\to text}(X)=D
\end{equation}

The function \(f_{Temp}^{Sig \to text}\) represents a transformation function that takes signal data as input and provides a description as output. Here, “\(Temp\)” signifies the template for the textual description, and \(D\) represents the final textual description. This approach helps create unified and precise textual representations, enhancing the interpretability and usability of signal data for large language models (LLMs). The text description templates are textual representations of the mathematical characteristics of signal data, providing both quantitative and qualitative descriptions of the mathematical features of composite signals. The steps and description content are illustrated in Figure 2. Specifically, the process begins with a qualitative description of the signal’s name and composition characteristics based on its time-domain waveform, identifying the signal units and operators. This is followed by a quantitative description of the signal’s time-domain characteristics, including its period, amplitude, and other indicators. Then, the signal’s frequency-domain characteristics are quantified, covering its fundamental frequency and amplitude, frequency-domain indicators, and the presence of harmonics, sidebands, and carrier frequencies. Finally, a comprehensive qualitative description connects the time-domain waveform characteristics with corresponding frequency-domain features. For simple harmonic signals, the text description templates of the five most common signal types can be referenced in Table 3.

\setlength{\LTleft}{0pt}
\setlength{\LTright}{0pt}

\begin{longtable}{|c|p{2cm}|p{10cm}|}
\caption{Signal Text Description Templates} \\
\hline
\textbf{No.} & \textbf{Type} & \textbf{Text Description} \\ \hline
\endfirsthead

\hline
\textbf{No.} & \textbf{Type} & \textbf{Text Description} \\ \hline
\endhead

\hline
\endfoot

\hline
\endlastfoot

1 & Single Harmonic Signal & This signal is a simple harmonic periodic signal. \\
  &                        & In the time-domain waveform of this signal, the period is $XX$ seconds, the amplitude is $YY$ mm/sec, and the phase is $B$ radians. \\
  &                        & In the spectrum of this signal, the frequency is $1/XX$ Hz, the amplitude of the frequency is $AA$ mm/sec, and the phase is $B$ radians. \\
  &                        & In the time-domain waveform of this signal, the waveform is a harmonic signal, so in the frequency spectrum, only a single frequency of $1/XX$ Hz can be observed. \\ \hline

2 & Multiple Harmonic Signal & This signal is a multi-harmonic periodic signal, that is, a non-simple harmonic periodic signal. \\
  &                          & In the time-domain waveform of this signal, the signal period is $XX$ seconds, and the amplitude is $YY$ mm/sec. \\
  &                          & In the frequency spectrum of this signal, the frequency of the fundamental (1st harmonic) is $ZZ$ Hz, amplitude is $AA$ mm/sec, phase is $B$ radians; the frequency of the 2nd harmonic is $ZZ2$ Hz, amplitude is $AA2$ mm/sec, phase is $B2$ radians; the frequency of the 3rd harmonic is $ZZ3$ Hz, amplitude is $AA3$ mm/sec, phase is $B3$ radians; the frequency of the 4th harmonic is $ZZ4$ Hz, amplitude is $AA4$ mm/sec, phase is $B4$ radians; $\ldots$; the frequency of the 8th harmonic is $ZZ8$ Hz, amplitude is $AA8$ mm/sec, phase is $B8$ radians; the frequency of the 9th harmonic is $ZZ9$ Hz, amplitude is $AA9$ mm/sec, phase is $B9$ radians; the frequency of the 10th harmonic is $ZZ10$ Hz, amplitude is $AA10$ mm/sec, phase is $B10$ radians. \\
  &                          & In the time-domain waveform of this signal, the waveform is distorted or asymmetric, that is, non-simple harmonic but periodic, so in the frequency spectrum, harmonics of $ZZ$ Hz can be observed. \\ \hline

3 & Random Harmonic Signal & This signal is a random harmonic signal, that is, a signal obtained by superimposing multiple random harmonic signals. \\
  &                        & In the time-domain waveform of this signal, the waveform is more complex and may be periodic or non-periodic, which needs to be determined based on the relationships between the random frequencies. \\
  &                        & In the frequency spectrum of this signal, the first frequency is $AA$ Hz, amplitude is $AA1$ mm/sec; the second frequency is $BB$ Hz, amplitude is $BB1$ mm/sec; the third frequency is $CC$ Hz, amplitude is $CC1$ mm/sec; the fourth frequency is $DD$ Hz, amplitude is $DD1$ mm/sec; the fifth frequency is $EE$ Hz, amplitude is $EE1$ mm/sec; the sixth frequency is $FF$ Hz, amplitude is $FF1$ mm/sec; the seventh frequency is $GG$ Hz, amplitude is $GG1$ mm/sec; $\ldots$; the $N$th frequency is $ZZ$ Hz, amplitude is $ZZ1$ mm/sec. \\
  &                        & This signal is a superposition of $N$ random harmonic components, so $N$ random frequencies can be observed in the frequency spectrum. \\ \hline

4 & Composite Harmonic Signal & This signal is a composite harmonic signal, obtained by superimposing multi-harmonic signals and random harmonic signals. \\
  &                           & In the frequency spectrum of this signal:

For the first multi-harmonic signal:
- Fundamental frequency: $ZZ$ Hz, amplitude: $AA$ mm/sec;
- 2nd harmonic: $ZZ2$ Hz, amplitude: $AA2$ mm/sec, phase: $B2$ radians;
- 3rd harmonic: $ZZ3$ Hz, amplitude: $AA3$ mm/sec, phase: $B3$ radians;
- $\ldots$
- 9th harmonic: $ZZ9$ Hz, amplitude: $AA9$ mm/sec, phase: $B9$ radians;
- 10th harmonic: $ZZ10$ Hz, amplitude: $AA10$ mm/sec, phase: $B10$ radians.

For the second multi-harmonic signal:
- Fundamental frequency: $MM$ Hz, amplitude: $NN$ mm/sec, phase: $P$ radians;
- $\ldots$
- 10th harmonic: $MM10$ Hz, amplitude: $NN10$ mm/sec, phase: $P10$ radians.

$\ldots$

For the $N$th multi-harmonic signal:
- $\ldots$

Random harmonic components:
- 1st frequency: $AA$ Hz, amplitude: $AA1$ mm/sec;
- 2nd frequency: $BB$ Hz, amplitude: $BB1$ mm/sec;
- 3rd frequency: $CC$ Hz, amplitude: $CC1$ mm/sec;
- $\ldots$
- $M$th frequency: $ZZ$ Hz, amplitude: $ZZ1$ mm/sec. \\
  &                           & This signal is a superposition of $N$ multi-harmonic signals and $M$ random harmonic components, so multiple fundamental harmonics and random frequencies can be observed in the frequency spectrum. \\ \hline

5 & Amplitude-Modulated Signal & This signal is an amplitude-modulated signal, where the carrier is a non-simple harmonic periodic signal, and the modulating wave is a simple harmonic signal. \\
  &                            & In the time-domain waveform of this signal, the carrier period is $T$ seconds, and the amplitude modulation period is $TT$ seconds. \\
  &                            & In the frequency spectrum of this signal:
- Carrier fundamental frequency: $FF$ Hz, amplitude: $AA$;
- Modulation frequency: $BB$ Hz;
- Sidebands appear on both sides of the carrier frequency $FF$ Hz at intervals of $BB$ Hz.
  - Left sidebands amplitudes (first 5): $XX$, $XX$, $XX$, $XX$, $XX$;
  - Right sidebands amplitudes (first 5): $XX$, $XX$, $XX$, $XX$, $XX$;
- Sidebands also appear around the second harmonic $2 \times FF$ Hz at intervals of $BB$ Hz.
  - Left sidebands amplitudes (first 5): $XX$, $XX$, $XX$, $XX$, $XX$;
  - Right sidebands amplitudes (first 5): $XX$, $XX$, $XX$, $XX$, $XX$;
- $\ldots$ \\
  &                            & The carrier waveform in the time domain is distorted and asymmetric, indicating a non-simple harmonic signal; therefore, harmonics at $FF$ Hz and sidebands at intervals of $BB$ Hz are visible in the spectrum. \\ \hline

\end{longtable}

\subsection{Sig2Txt Based on Pre-trained Large Models}

In this framework, a large multimodal model is selected as the foundational model, comprising a data encoder (such as CLIP), a connector, and a large language model. Both the encoder and language model are fine-tuned using LoRA. The model is trained and optimized to enhance its performance in signal parameter recognition tasks by using "signal data (e.g., time series and image data) and their mathematical feature-text description" Q\&A pairs, denoted as \( G = (X_s, X_{q_i}, X_{a_i}) \), constructed from a signal generator module or real signals.

Two approaches can be taken for constructing the Q\&A pairs in the model's dataset. The first is through simulated signal data generated automatically by a custom-built signal generator module. The second approach utilizes real signal data. When using data generated by the signal generator module, signal denoising algorithms are generally not needed in subsequent processing. However, when using real monitoring data, the signal denoising module proposed in this paper must be applied. During the Sig2Txt application phase, time series data typically undergo denoising using the proposed signal denoising algorithm before being input into Sig2Txt to improve model accuracy.

To generate Q\&A pairs more efficiently, a signal generator module has been developed in this section. Figure 3 illustrates the framework for generating "signal data-mathematical feature-text description" Q\&A pairs using the signal generator module. This framework primarily includes six components: signal unit, signal operator, signal function, random parameters, composite signal, and text description.

\begin{figure}[H]
    \centering
    \includegraphics[width=1\linewidth]{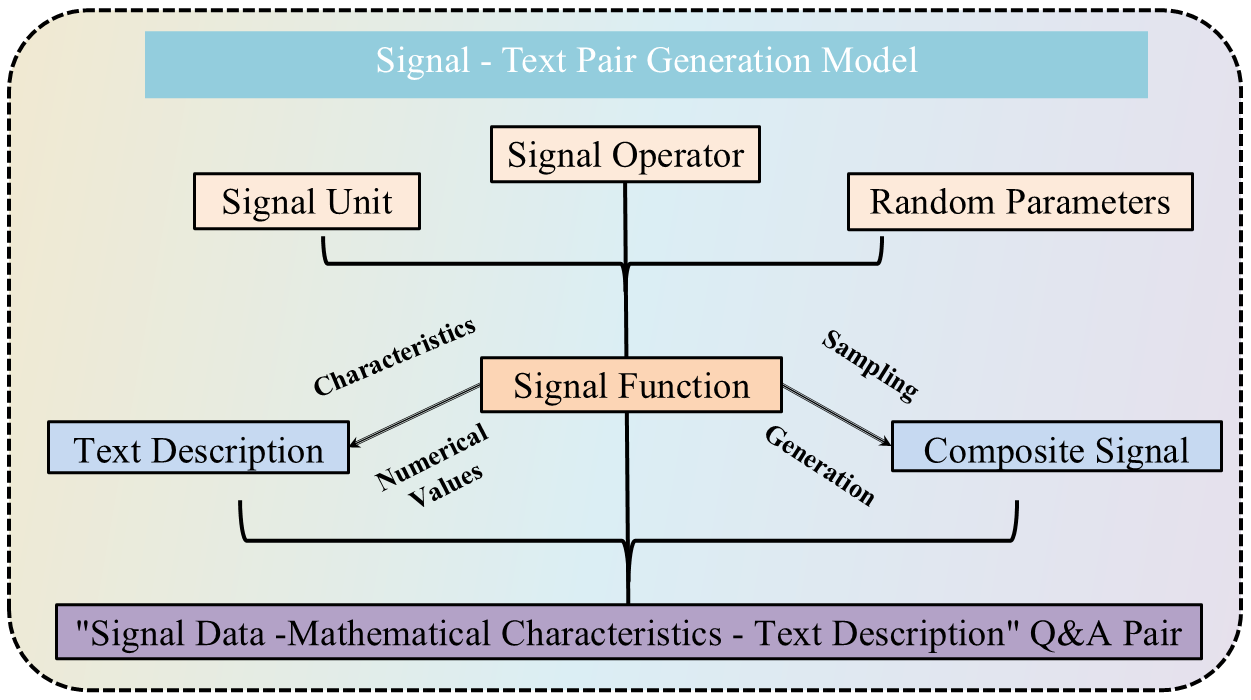}
    \caption{`Signal Data - Mathematical Features - Text Description' Question-Answer Pair Generation Framework}
    \label{fig:enter-label}
\end{figure}

(1) Construct Signal Unit
A signal unit refers to the basic types of signals that constitute a signal function, including simple harmonic signals, impulse decay signals, wavelet signals, non-stationary random signals, and stationary random signals. Among these, a simple harmonic signal is represented as:

\begin{equation}
x_{sh} = A \sin(2 \pi f t + \varphi)
\end{equation}

Where \( \varphi \) is the initial phase, \( A \) is the amplitude of the signal, \( f \) is the frequency, and \( t \) is time.

The impulse decay signal is represented as:
\begin{equation}
x_{sa} = A e^{-\alpha t} \sin(2 \pi f t + \varphi)
\end{equation}
Where \( \varphi \) is the initial phase, \( A \) is the impulse amplitude of the signal, \( f \) is the frequency, \( t \) is time, and \( \alpha \) is the decay time constant.

The wavelet signal is represented as:
\begin{equation}
x_{sw} = \frac{1}{\sqrt{a}} \psi \left( \frac{t - b}{a} \right)
\end{equation}
Where \( a \) is the scaling parameter, \( b \) is the time shift parameter, and \( \psi(t) \) is the mother wavelet, also known as the basic wavelet.

A random signal refers to a time series whose values at each moment are random variables. Although the values of a random signal cannot be determined in advance, they generally follow certain statistical laws. In other words, random signals can be described in terms of statistical properties through probability distributions. Based on whether the \( k \)-th order moment is related to time, random signals can be divided into non-stationary random signals and stationary random signals. The \( n \)-th order stationary random signal \( x_{ws} \) is defined as:
\begin{equation}
\mu(t_1, \cdots, t_k) = \mu(t_1 + \tau, \cdots, t_k + \tau)
\end{equation}
Where \( \mu(t) \) represents the mean value of the stationary random signal.

(2) Construct Signal Operator
A signal operator refers to the method of performing operations between signal units, including addition, subtraction, multiplication, division, convolution, accumulation, and integration. Specifically, the convolution formula for signal unit \( f(x) \) and signal unit \( h(x) \) is as follows:
\begin{equation}
\mu(t_1, \cdots, t_k) = \mu(t_1 + \tau, \cdots, t_k + \tau)
\end{equation}

The accumulation of the signal unit \( f(x) \), represented as \( f(x)^n \), is as follows:
\begin{equation}
f(x)^n = \underbrace{f(x) f(x) \cdots f(x)}_{n}
\end{equation}

The integral of the signal unit \( f(x) \), denoted as \( \mu(x) \), is as follows:
\begin{equation}
\mu(x) = \int_a^b f(x) \, dx
\end{equation}

(3) Construct Random Parameters
Random parameters refer to the variable parameters within signal units and signal operators. The variable parameters in signal units include: the amplitude, frequency, and phase in simple harmonic signals; the impulse amplitude, frequency, and decay time constant in impulse decay signals; the scaling parameter and time shift parameter in wavelet signals. The variable parameters in signal operators include adjusting the number or ratio of addition, subtraction, multiplication, division, convolution, power operation, and integration operations, as well as the number or proportional coefficients of signal operators.

(4) Construct Signal Function
A signal function refers to a mathematical expression composed of signal units, signal operators, and random parameters. It contains all the quantitative parameters needed to describe a signal. Below are the function formulas for several typical signals:
Multiple Harmonic Signals:
\begin{equation}
y_1(t) = \sum_{i=1}^N A_i \sin(2 \pi f t + \varphi_i)
\end{equation}
Where \( \varphi_i \) is the initial phase, \( A_i \) is the amplitude of the signal, \( f \) is the frequency, \( t \) is time, and \( N \) is the total number of harmonics.
Multiple Random Harmonic Signals:
\begin{equation}
y_2(t) = \sum_{i=1}^N B_i \sin(2 \pi f_i t + \varphi_i)
\end{equation}
Where \( \varphi_i \) is the initial phase, \( B_i \) is the amplitude of the signal, \( f_i \) is the frequency, \( t \) is time, and \( N \) is the total number of harmonics.
Multiple harmonic signals combined with random harmonic signals:
\begin{equation}
y(t) = y_1 + y_2
\end{equation}
Bearing vibration signal function:
\begin{equation}
y(t) = A_0 e^{-\omega (t - k * 1 / f_d)} \cos[2 \pi f_n (t - k * 1 / f_d)] u(t - k * \frac{1}{f_d} - k_1 * \frac{1}{f_d})
\end{equation}
Where \( A_0 \) is the amplitude, \( u(t) \) is the unit step function, \( f_n \) is the system natural frequency, \( f_d \) is the characteristic frequency of the fault, \( k \) is an integer, \( k_1 \) is either 0 or the value 1/2. For outer and inner ring faults, \( k_1 = 0 \); for rolling element faults, \( k_1 = 1/2 \).
Gear vibration signal function:
\begin{equation}
y(t) = \sum_i^N \left[ (1 + A_i \cos(2 \pi f_{ch} t) + \psi_i) \cos(2 \pi k f_m t) + B_i \sin(2 \pi f_{ch} t + \alpha_i) + \theta_i \right]
\end{equation}
Where \( \psi_i \), \( \alpha_i \), and \( \theta_i \) are the phases, \( A_i \) and \( B_i \) are the amplitudes of the signal, \( f_m \) and \( f_{ch} \) are the fault characteristic frequency and meshing frequency, \( t \) is time, and \( N \) is the maximum order of amplitude modulation and frequency modulation.

Next, the data generated by the aforementioned signal functions is stored as time-series data to facilitate the subsequent signal-to-text conversion process. This step ensures that the generated signals are properly structured and ready for analysis, providing a foundation for extracting meaningful textual descriptions from the signal characteristics.

\begin{figure}[H]
    \centering
    \includegraphics[width=1\linewidth]{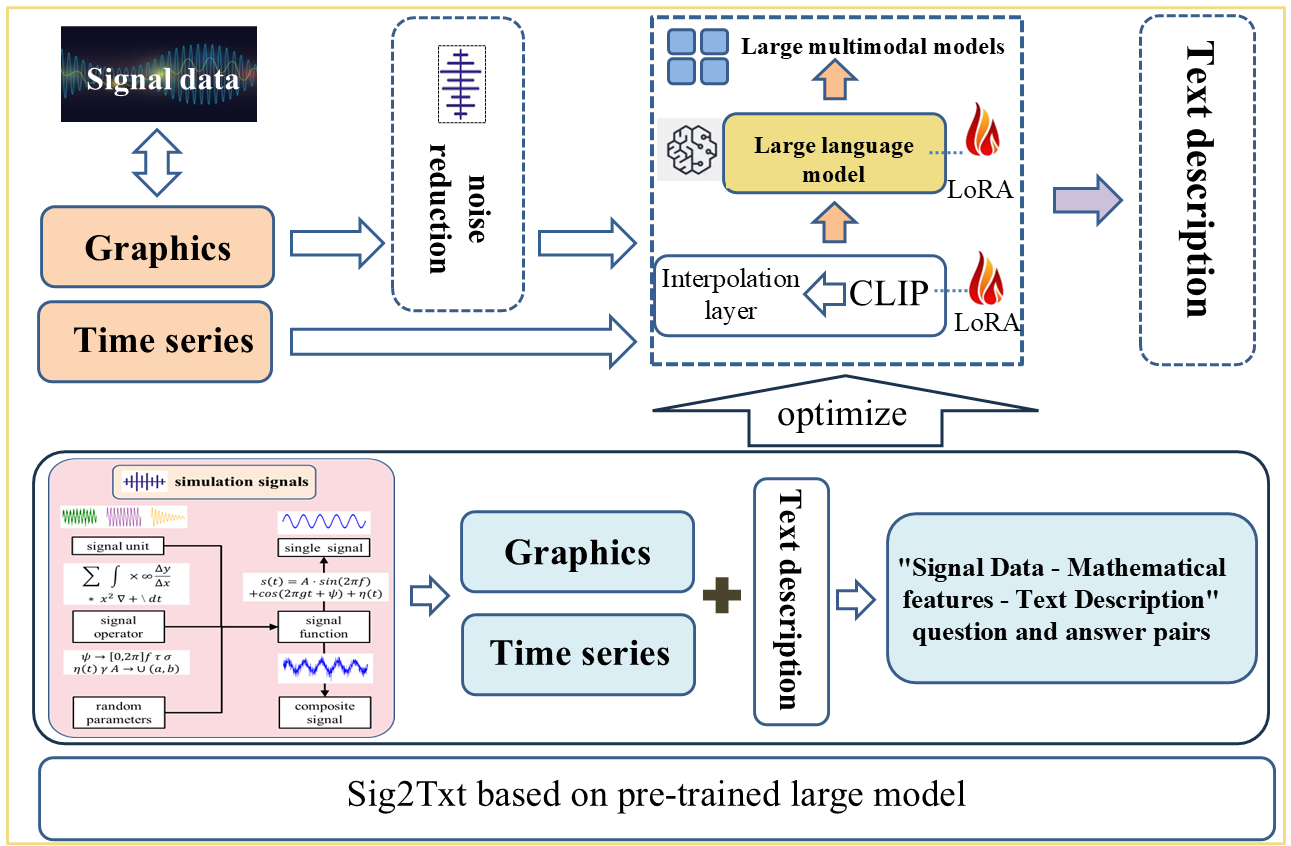}
    \caption{The Flowchat For Pretrained Sig2Txt-Based Model}
    \label{fig:enter-label}
\end{figure}

\section{The Professional Text Model for Intelligent Operations and Maintenance}

\subsection{Text Dataset}

To ensure the training data for large text models in the intelligent diagnostics domain is both broad and specialized, high-quality text data was gathered from multiple sources. These sources include academic papers, textbooks, international standards, as well as training and certification materials for vibration analysts.

To build a guided dataset for the model, a systematic approach was adopted to manually screen and extract information related to fault diagnosis of rotating machinery from numerous academic references and standard textbooks. This extracted information was then organized and restructured into a "uniformly formatted" and "consistently structured" corpus, presented in a single-turn question-answer dialogue format. For example, the research generated a series of dialogues that contain questions and answers about fault characteristics, comprehensive analysis procedures for fault pattern recognition, and detailed fault repair steps. This method effectively consolidated a structured and focused data resource that distills the core elements of fault diagnosis and resolution strategies, providing a solid foundation for subsequent research.

During the data preprocessing phase, a series of advanced natural language processing techniques and tools were adopted to ensure the information extracted from the raw text could effectively meet model training requirements. Using Python's PyMuPDF library, text content was extracted from collected PDF documents. Known for its efficient PDF parsing capability, the PyMuPDF library accurately extracts text, images, and tables. Several challenges were encountered due to inconsistencies in PDF formatting, which were effectively addressed by optimizing parsing parameters, implementing preprocessing steps, and applying Optical Character Recognition (OCR) technology.

To facilitate subsequent processing, the extracted text was segmented and labeled. The SpaCy natural language processing library was used for precise segmentation of the text, dividing it into sentences and paragraphs and performing basic part-of-speech tagging and dependency analysis. SpaCy's high-efficiency tokenization and tagging capabilities provided a solid foundation for generating question-answer pairs.

For dataset construction to convert text information into question-answer pairs, the Bonito library was utilized. This library offers diverse tools for generating and optimizing question-answer pairs, thus promoting the process of creating a high-quality question-answer dataset tailored to specific needs. During dataset construction, the generated question-answer pairs underwent detailed manual verification and optimization to ensure accuracy and professionalism in the dataset.

Through this process, a high-quality dataset with broad and specialized content was successfully constructed, providing a reliable foundation for training large-scale text models in the intelligent diagnostics field.

\subsection{Model}
In the field of intelligent operations and maintenance, the overall design approach for large text models is rooted in meticulous fine-tuning of a general-purpose foundational model. This process leverages a specifically curated domain-specific dataset to enhance the model’s capability in addressing intelligent maintenance and operations queries. The ultimate goal is for the fine-tuned professional text model to achieve or even surpass the expertise ISO Level III vibration analyst, thereby providing engineers with efficient and precise intelligent maintenance support.

Figure 4 illustrates the operational framework of the professional text model, covering all stages from receiving user input to generating responses. In the initial step, users submit queries in text form about issues encountered during the routine operation of mechanical equipment via a client interface. The submitted text is then processed through stages of preprocessing, tokenization, and word embedding to ensure that the content is optimally parsed by the model. Using the Q-Former encoding mechanism, the text information is converted into high-dimensional data representations rich in semantic and contextual information, creating a data input format suitable for the model. Finally, the preprocessed query data is fed into the finely-tuned model, which outputs professional and well-reasoned answers for user reference.

\begin{figure}[H]
\centerline{\includegraphics[width=0.9\linewidth]{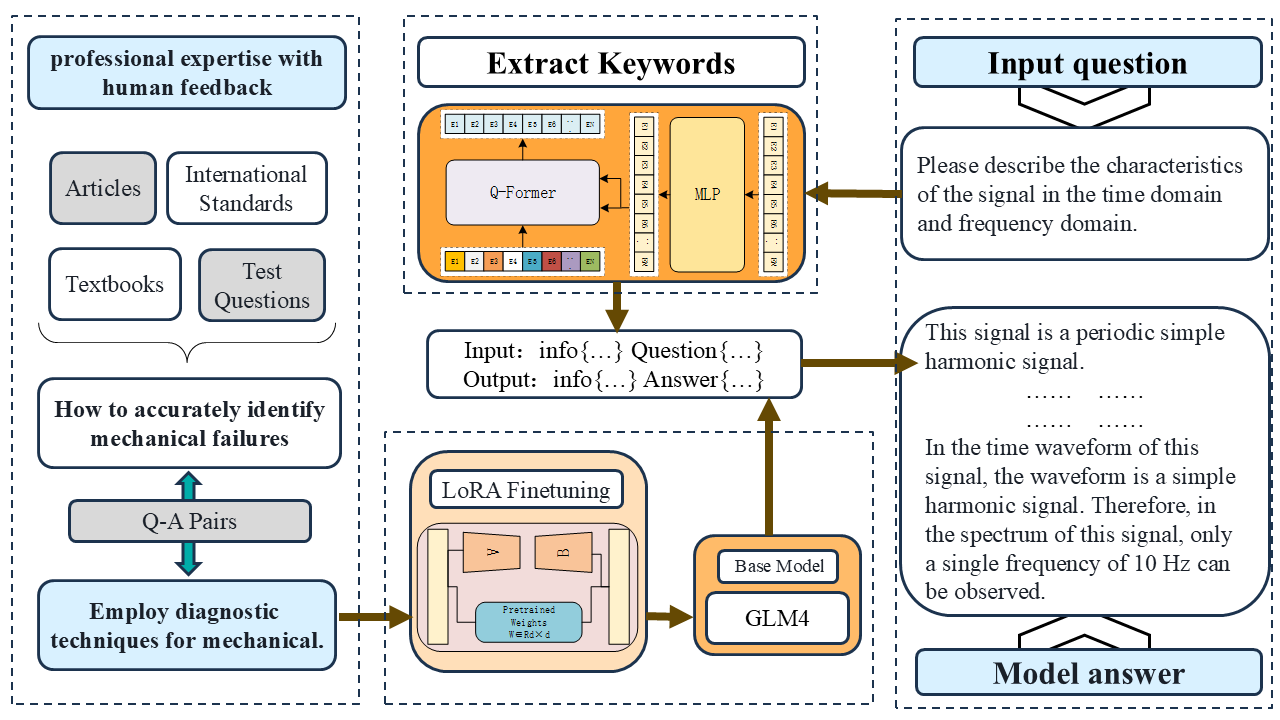}}
\caption{The flowchart of the text-based large-scale model for intelligent maintenance.}
\label{fig framework}
\end{figure}

For the development of this model, over 5,000 high-standard guidance data entries were gathered specifically for the supervised fine-tuning process. Model development is based on the LLaMA-Factory open-source framework. All foundational language models employ a decoder-only architecture, achieved through autoregressive language modeling techniques. The base model adopted for this study is GLM4, renowned for its efficiency in autoregressive language modeling. GLM4 provides multiple parameter-scaled versions to support diverse application scenarios, including versions with 9B and 130B parameters. The 9B parameter model version was selected for this study. 

The specific configuration parameters of the chosen model are as follows: the model size is 9B, with 36 layers, each having a hidden dimension of 4096, 32 attention heads, and a maximum context length of up to 128,000 tokens. With its impressive parameter efficiency and text generation capabilities, the GLM4 model demonstrates exceptional performance across a wide range of natural language processing tasks.

The central concept of this method is that the joint probability of tokens in a text can be expressed as in Eq. 20.

\begin{equation}
    p(w) = p(w_1, w_2, \ldots, w_t) = \prod_{t=1}^{T} p(w_t | w_{<t})
\end{equation}

 Here, \(w\) represents a sequence of tokens, \(w_t\) is the \(t^{th}\) token, and \(w_{<t}\) is the sequence of tokens preceding \(w_t\).

 To enhance model training efficiency, this study adopted the Ubuntu operating system, widely used in deep learning tasks due to its stability. On the hardware side, an Intel Xeon Platinum 8352V 36-core processor, known for its powerful computing and multi-threading capabilities, was selected to effectively support the complex training requirements of large-scale data processing. For memory, 256GB of RAM was configured to ensure stable and efficient system performance when handling large datasets and conducting long-duration training tasks. Additionally, to further accelerate model training and improve training results, the setup includes eight NVIDIA Geforce RTX4090 GPUs. These GPUs offer exceptional floating-point performance and support high-efficiency, large-scale parallel computing, significantly reducing the model training period. Detailed training configuration information is shown in Table 4.

For the software framework, this study employed PyTorch 2.1, a powerful and highly flexible deep learning framework that supports the construction, training, and deployment of various deep learning models. The convenience and efficiency of PyTorch 2.1 facilitate model fine-tuning and optimization, while its extensive libraries and tool resources help streamline the development process.

\begin{table}[H]
    \centering
    \caption{Experimental Hardware Conditions}
    \begin{tabular}{l l}
    \hline
         Hardware Name & Specification\\ \hline
         Operating System & Ubuntu  \\
         Intel CPU & Intel Xeon Platinum 8352V 36-core Processor \\
         Memory & 256GB \\
         Framework & Pytorch 2.1 \\
         GPUs & 8×NVIDIA Geforce RTX4090 \\ \hline
    \end{tabular}

    \label{tab:my_label}
\end{table}

\subsection{Training Process}
In the initial phase, this study built a specialized knowledge base by performing in-depth analysis and information extraction from academic papers, textbooks, international standards, and vibration analyst training materials. The PyMuPDF library in Python was used for PDF text extraction, followed by text segmentation processing with the SpaCy library, and then Bonito was utilized to generate question-answer pairs. These steps ensured high-quality standards and consistency in the dataset, allowing the model to cover a wide range of topics within the intelligent operations and maintenance domain.

To adapt the GLM4 foundational model to the specific requirements of the intelligent maintenance domain, this study applied Low-Rank Adaptation (LoRA) for model fine-tuning. This technique reduces the dimensionality of parameter updates by introducing low-rank matrices, aiming to decrease computational costs and storage needs while maintaining or enhancing model performance. Specifically, the pre-trained weight matrix \( W_o \in \mathbb{R}^{m \times n} \) undergoes low-rank decomposition:
\begin{equation}
W = W_o + \Delta W
\end{equation}

Among them, \( \Delta W = BA \), \( B \) and \( A \) are two trained parameter matrices. The dimension of \( B \) is \( m \times r \), and the dimension of \( A \) is \( r \times n \), where \( r \ll \min(m, n) \) is the rank. \( W \) is the new weight matrix. Therefore, it is not necessary to train all parameters again, and only the weights in \( A \) and \( B \) are updated using gradient descent. \( W_o \) and the injected matrix \( \Delta W \) are both multiplied by the input \( X \), and the coordinates are then combined to obtain the feature representation suitable for the new task. Thus, this new forward propagation calculation is:

\begin{equation}
F = WX = W_o X + \Delta W X = W_o X + BA X
\end{equation}

In this equation, \( F \) represents the output features, and \( X \) represents the input features. \( W_o X \) is the original output without LoRA. \( A \) is initialized using random Gaussian initialization, and \( B \) is set to zero to ensure \( \Delta W = 0 \) at the beginning stage. Additionally, in practice, \( \Delta W X \) is scaled by \( \alpha / r \) to stabilize the learning of injected parameters, where \( \alpha \) is a hyperparameter.

Table 5 presents the parameter settings for model training, which ensure stability and efficiency throughout the training process. During training, the loss value is regularly monitored and recorded, with the learning rate dynamically adjusted based on changes in the loss value. A cosine annealing learning rate scheduling strategy is applied to optimize model performance at each stage of training.

\begin{table}[H]
\centering
\caption{Training Parameter Settings}
\begin{tabular}{l c}
\hline
\textbf{Parameter} & \textbf{Instruction Fine-Tuning Parameter} \\ \hline
Cutoff\_len        & 1024                                      \\ 
Learning\_rate     & 5.0e-05                                   \\ 
LoRA\_alpha        & 16                                        \\ 
LoRA\_dropout      & 0                                         \\ 
LoRA\_rank         & 16                                        \\ 
Num\_train\_epochs & 20                                        \\ \hline
\end{tabular}
\end{table}

\section{Generative Multimodal Large Model (MaintAGT)}
\subsection{Multimodal Large Model Architecture}
Figure 6 illustrates the generative multimodal large model architecture (MaintAGT) proposed in this study. The primary innovation of this model is the integration of a Signal-to-Text (Sig2Txt) module, which converts monitoring data, such as vibration signals, into textual descriptions and processes them alongside traditional text data within the large language model. This approach enables the model to simultaneously interpret and analyze data from different modalities, achieving comprehensive perception and diagnosis of equipment conditions.

\begin{figure}
    \centering
    \includegraphics[width=1\linewidth]{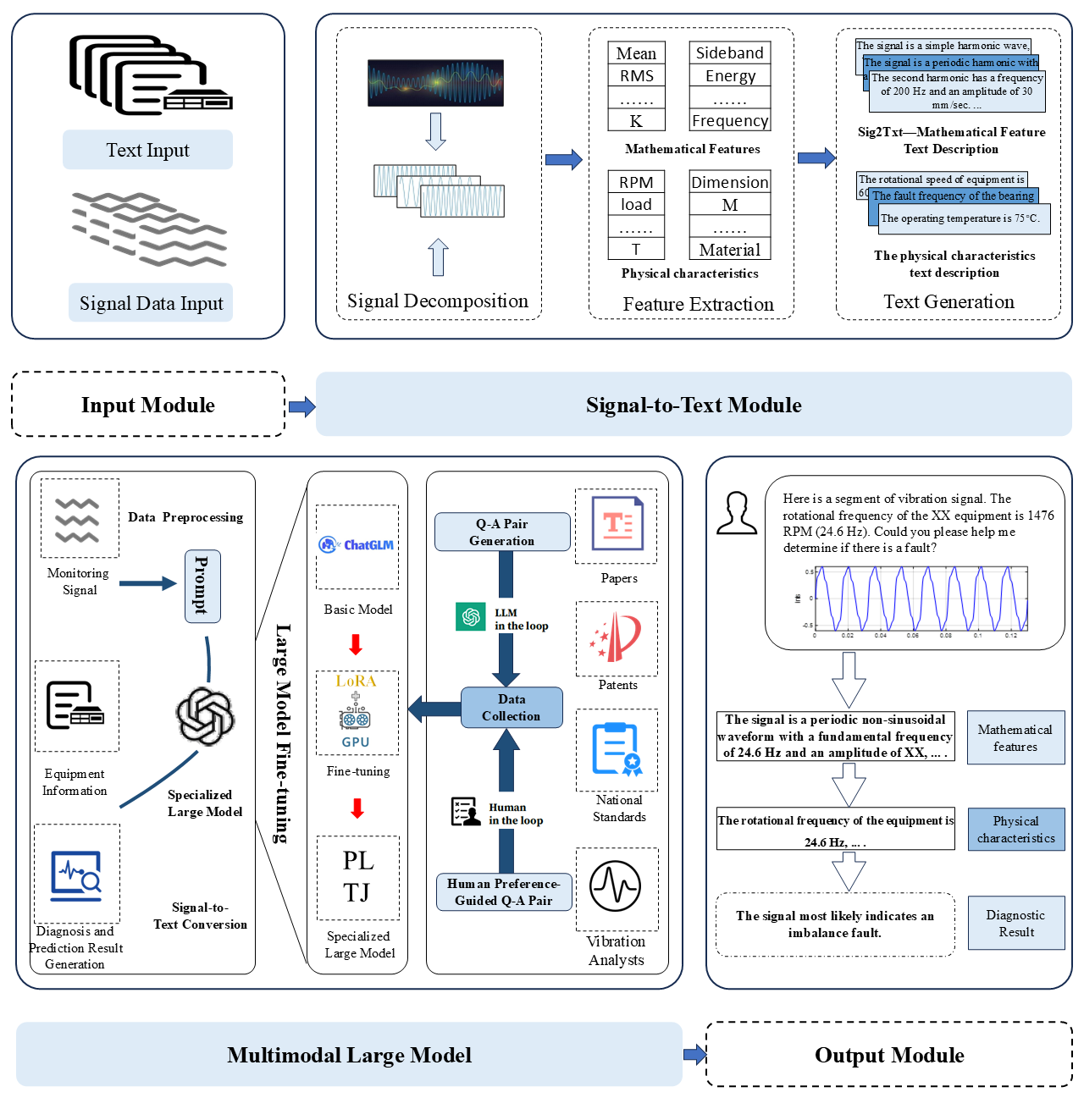}
    \caption{Intelligent Maintenance Multimodal Fusion Architecture}
    \label{fig:enter-label}
\end{figure}

In the modality fusion process, multimodal data must first undergo preprocessing. The Sig2Txt module employs a signal denoising algorithm and text conversion to transform raw signals into structured text descriptions, capturing both the time-domain and frequency-domain mathematical features of the signals. For conventional text data, such as equipment manuals, maintenance records, and fault reports, standard natural language processing is applied directly. The signal and text data are then input into the LoRA fine-tuned model, where the attention mechanism is utilized to fuse this data, capturing correlations and complementary information, and generating a comprehensive knowledge representation for intelligent diagnostics and question-answer pair generation. To ensure the accuracy and reliability of the model's diagnostic results, a "Human in the Loop" mechanism is presented, continuously optimizing model performance through human guidance and correction, thereby providing efficient and precise decision support for intelligent maintenance.

\subsection{Chain of Thought Architecture}
Building on the multimodal large model architecture, this section proposes a Chain of Thought reasoning mechanism, enabling the model to better capture and understand complex features of equipment operational states through step-by-step reasoning, thereby enhancing the accuracy and effectiveness of fault diagnosis.

Currently, fault diagnosis typically relies on a direct mapping between monitoring signals and fault patterns, following traditional methods that struggle to achieve in-depth analysis and understanding of signal features and equipment's physical information. To address this limitation, this study proposes a new paradigm that integrates the Chain of Thought approach. Figure 6 presents a framework comparison between the traditional fault diagnosis paradigm and the new paradigm proposed in this study. By incorporating Chain of Thought reasoning, the new paradigm facilitates a step-by-step analysis of input data, allowing the model to explore complex relationships between signal features and physical information, achieving more refined feature extraction and analysis.

\begin{figure}[H]
    \centering
    \includegraphics[width=1\linewidth]{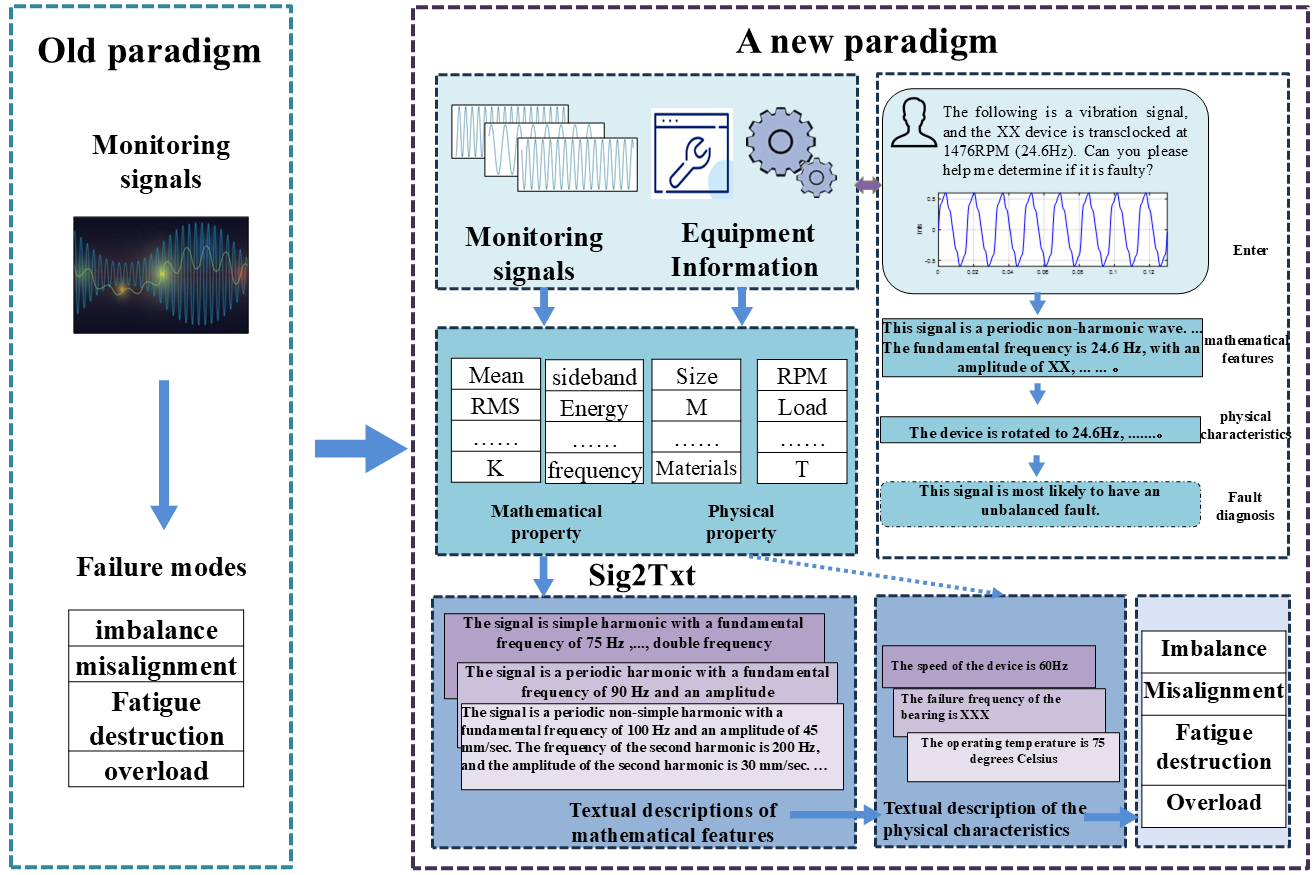}
    \caption{Comparison of Traditional and New Paradigm Frameworks for Fault Diagnosis}
    \label{fig:enter-label}
\end{figure}

Chain of Thought (CoT) is a technique designed to enhance the reasoning capabilities of large language models by simulating the step-by-step reasoning and analysis processes of human experts. This enables the model to perform multi-step thinking and reasoning in complex tasks, allowing it to assess equipment operational states, identify potential faults, and provide reasonable maintenance recommendations. In contrast, traditional paradigms lack in-depth processing and comprehensive analysis of monitoring data and equipment information, often mapping monitoring signals directly to fault patterns without deep exploration of signal features or thorough consideration of physical characteristics. The new diagnostic paradigm, however, offers powerful reasoning and analytical capabilities.

As illustrated in Figure 7, the multimodal Chain of Thought model framework primarily involves several steps. First, the model converts input monitoring signals into mathematical feature descriptions via the Sig2Txt module, which is based on prior algorithmic knowledge, thus obtaining key mathematical features of the signal data. Next, the model retrieves textual descriptions of the equipment's physical characteristics from the equipment information. Combining the mathematical features with the physical feature descriptions, the model infers the probable fault state of the equipment. This comprehensive analytical approach, based on the mathematical features of signals and physical information of the equipment, enhances the accuracy and intelligence of fault diagnosis, providing a more reliable basis for assessing the health of complex systems. 

\subsection{Training Configuration}
To establish a closer connection between the model architecture and reasoning mechanism, this study presented optimized adjustments at the training configuration level. Building on the hardware and software settings detailed in Section 2.3, further refinements were made to data preprocessing and model training specifics. A flexible tensor singular spectrum decomposition method was applied for denoising real-world noisy signals, ensuring that critical feature extraction remains unaffected by noise. The Signal-to-Text (Sig2Txt) module was utilized to extract the mathematical characteristics of monitoring signals in both the time and frequency domains, outputting standardized text descriptions that accurately reflect the primary features of the signals.

The generated signal text, combined with other text data, was then input into the base language model GLM4. Through a supervised learning strategy, the model acquired the ability to understand and process mechanical fault signal data. In the training phase, distributed training techniques and a GPU cluster were employed to accelerate computation, facilitating faster parameter updates and overall training efficiency. The LoRA fine-tuned language model effectively enhances feature representation while ensuring computational efficiency, enabling a robust mapping between input data, equipment health status, and fault types.

Through these steps, the multimodal large model effectively integrates both text and signal data modalities, significantly improving diagnostic and predictive accuracy and reliability. This innovative approach contributes a new technological pathway and solution strategy for the field of intelligent maintenance.

\section{Experiments and Evaluation}
\subsection{Signal-to-Text Module Evaluation}
(1) Example 1: Single Harmonic Signal  
To thoroughly analyze the performance of the Signal-to-Text module, this study initially employed a set of specific simulated signals for evaluation. The characteristic parameters of this signal are detailed as follows:  
The signal frequency is set to 30 Hz, with an amplitude of 4 volts, a period of 33.33 milliseconds, and an initial phase of 0 degrees. The waveform is shown in Figure 8.

\begin{figure}[H]
    \centering
    \includegraphics[width=1\linewidth]{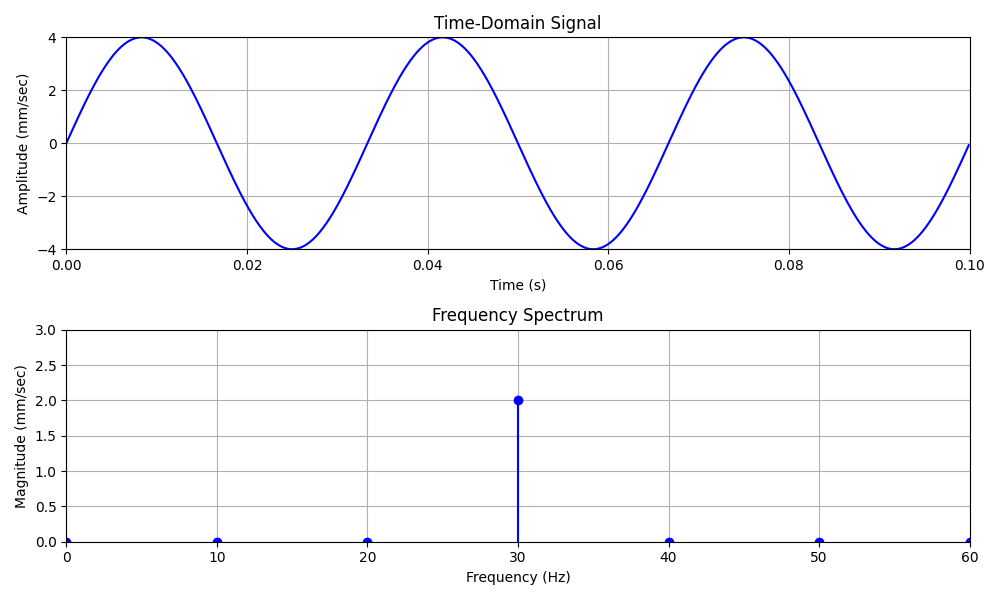}
    \caption{Simulated signal image used as input for the signal-to-text model}
    \label{fig:enter-label}
\end{figure}

In this study, a signal generator module developed in the MATLAB environment was used to construct a simulated signal and generate its corresponding signal image. The signal image was input into the signal-to-text model. The resulting text description is shown in Figure 9.





\begin{figure}
    \centering
    \includegraphics[width=1.1\linewidth]{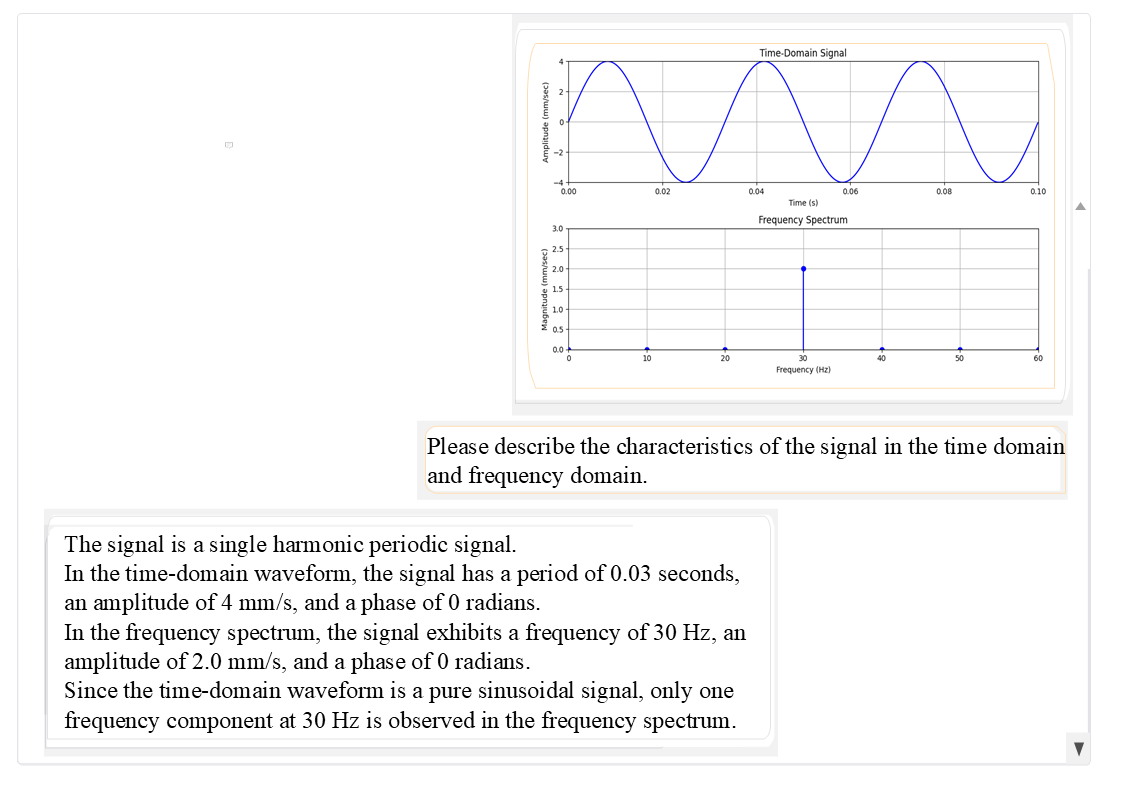}
    \caption{extual description of signal images}
    \label{fig:enter-label}
\end{figure}

(2) Example 2: Multi-Harmonic Signal  
Another signal generated by the signal generator is shown in Figure 9. After importing this signal into the Signal-to-Text module, the resulting text description is displayed in Table 7.

\begin{figure}[H]
    \centering
    \includegraphics[width=1\linewidth]{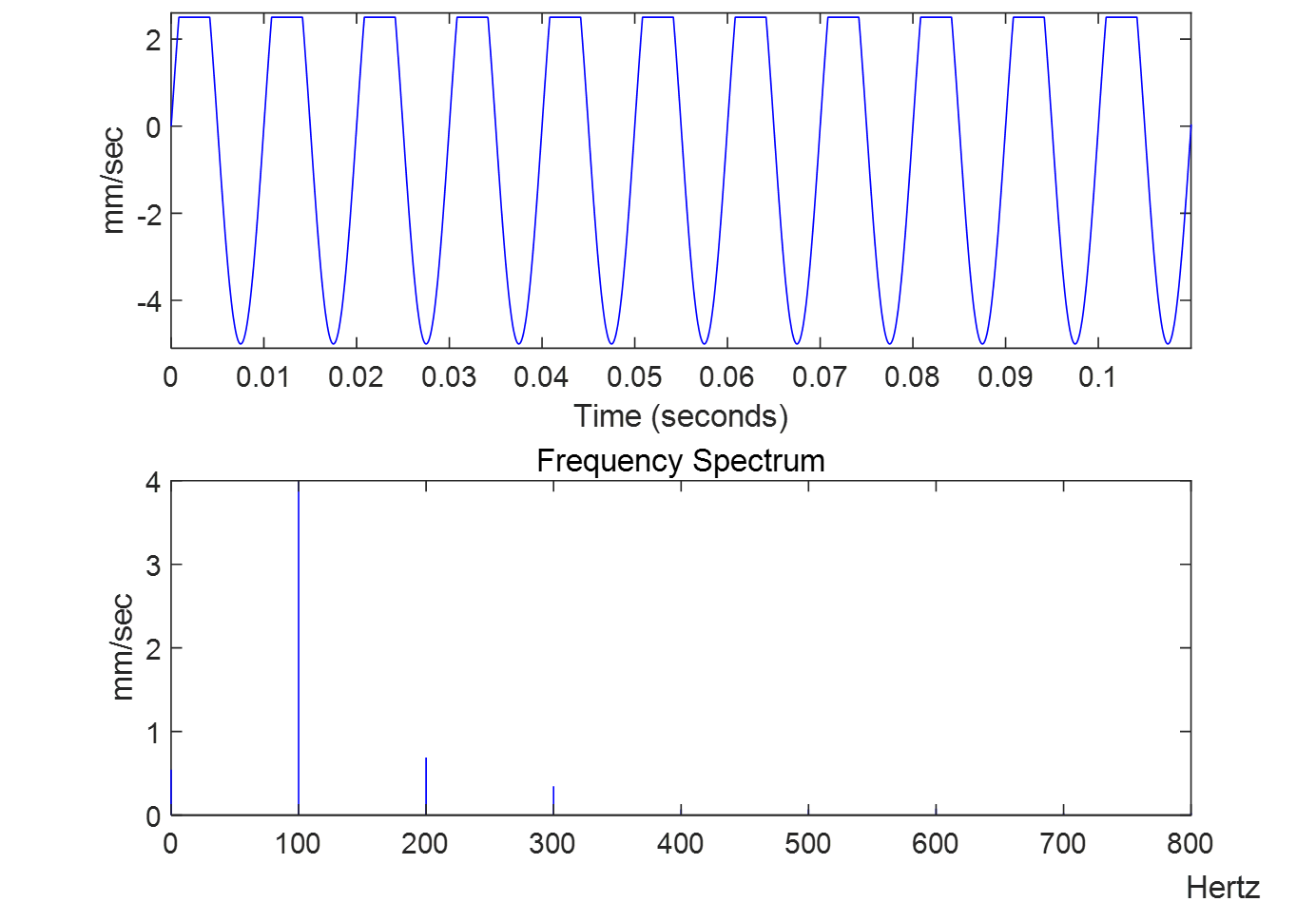}
    \caption{Input multi-harmonic vibration signals}
    \label{fig:enter-label}
\end{figure}

\begin{table}[H]
\centering
\caption{Text Description of multi-harmonic vibration signals}
\begin{tabular}{p{0.9\textwidth}}
\hline
\textbf{Text Description} \\ \hline
This signal is periodic but not a Simple Harmonic Signal.\\
In the time waveform of this signal, the period is 0.01 seconds. \\
In the spectrum of this signal, the base frequency is 100Hz, the amplitude of the base frequency is 4.02 mm/sec; the second harmonic frequency is 200Hz, the amplitude of the second harmonic frequency is 0.689 mm/sec; the third harmonic frequency is 300Hz, the amplitude of the third harmonic frequency is 0.344 mm/sec. \\ \hline
\end{tabular}
\end{table}

Through comparative analysis between the raw signals and the text descriptions, results indicate that the Signal-to-Text module demonstrates high accuracy in capturing critical features of real signals, including frequency, amplitude, and phase. However, in terms of describing the signal’s noise components and nonlinear characteristics, there remains room for improvement in the module’s representational capability. Thus, while the Signal-to-Text module shows significant accuracy and practicality in real signal analysis, further performance optimization is needed for handling complex signal structures.

Overall, the testing effectiveness with simulated signals indicates that the Signal-to-Text module performs excellently across various signal types, generating precise and detailed text descriptions. This technology not only provides a reliable data foundation for advanced diagnostics in intelligent maintenance systems but also confirms the module’s applicability and effectiveness in practical operational scenarios.

\subsection{Text Model Evaluation}
To evaluate the performance of the Chain of Thought paradigm-based multimodal large model (MaintAGT), this study conducted several comparative experiments, benchmarking it against industry-recognized advanced large models. The selected models include GPT, ChatGLM, and Qwen. Test questions designed for Category III vibration analysts were input into the constructed model, covering eight key knowledge areas in intelligent maintenance: Condition Monitoring, Vibration Principles, Data Acquisition, Signal Processing, Fault Analysis, Corrective Actions, Equipment Detection and Diagnosis, and Equipment Knowledge. The test results for each model are shown in Figure 8. 

Results indicate that the multimodal large model, MaintAGT, achieved an overall accuracy of 70\% in comprehensive testing, reaching the ISO Level III of a human Category vibration analyst. Furthermore, MaintAGT outperformed other models in all eight knowledge areas, demonstrating comprehensive and stable performance. MaintAGT showed balanced performance across all knowledge areas without any specific area excelling or underperforming relative to others. In contrast, the accuracy rates of GPT4o, ChatGLM4, and Qwen2.5 were lower than those of the multimodal large model MaintAGT proposed in this study. Specifically, GLM4 had the lowest accuracy in Fault Analysis, GPT4o had the lowest in Equipment Knowledge, while Qwen’s overall performance was slightly better than the other two models.

\begin{figure*}[htbp]
    \centering
    \begin{subfigure}[b]{1.0\linewidth} 
        \centering
        \includegraphics[width=1.0\linewidth, height=0.45\linewidth]{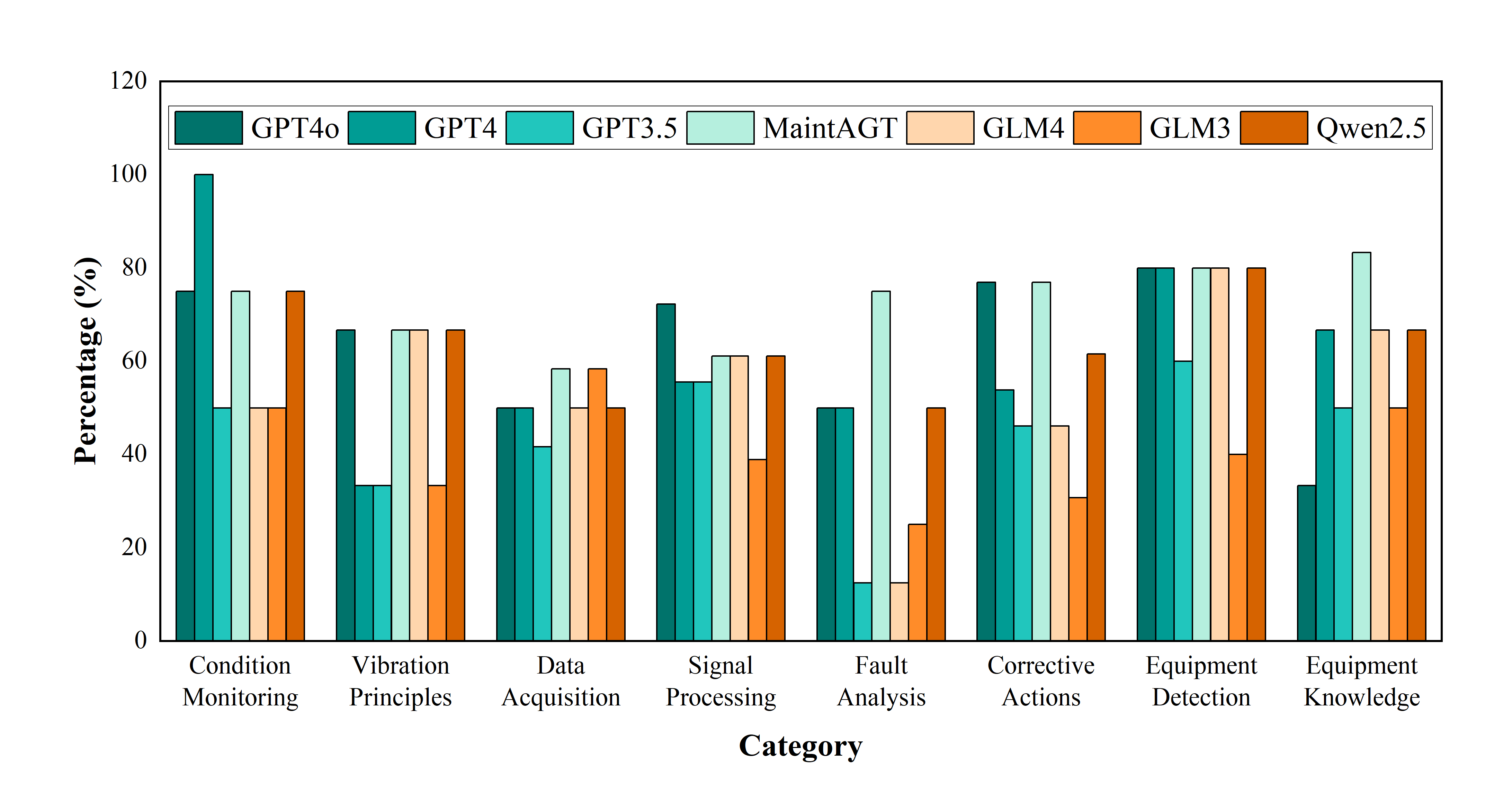}
        \caption{Comparison of MaintAGT with Other General Large Models.} 
    \end{subfigure}
    
    \vspace{0.5cm} 

    \begin{subfigure}[b]{0.9\linewidth} 
        \centering
        \includegraphics[width=1.0\linewidth, height=0.45\linewidth]{figs/Figure_11_B.png}
        \caption{Comparison on Datasets from Eight Different Domains.} 
    \end{subfigure}

    \caption{Comparison of Different Models on Various Categories.}
    \label{fig:framework}
\end{figure*}

\subsection{Multimodal Large Model Evaluation}
To comprehensively evaluate the multimodal data processing capabilities of the large model MaintAGT, a graphical-textual question from a vibration analyst training manual was selected as Example 3 for testing. This question includes signal data along with a question-answer segment for fault type identification of the signal, as shown in Figure 9.

\begin{tcolorbox}[colback=white, colframe=black, title=Evaluation results of multimodal large models., breakable, width=14cm]
    \begin{minipage}{0.5\textwidth}
        The rotational frequency of the equipment is 60 Hz. What type of fault or condition is the equipment experiencing?\\

        A. Unbalance\\
        
        B. Misalignment\\
        
        C. Looseness\\
        
        D. Blade Pass\\
    \end{minipage}%
    \begin{minipage}{0.5\textwidth}
        \centering 
        \includegraphics[width=\textwidth]{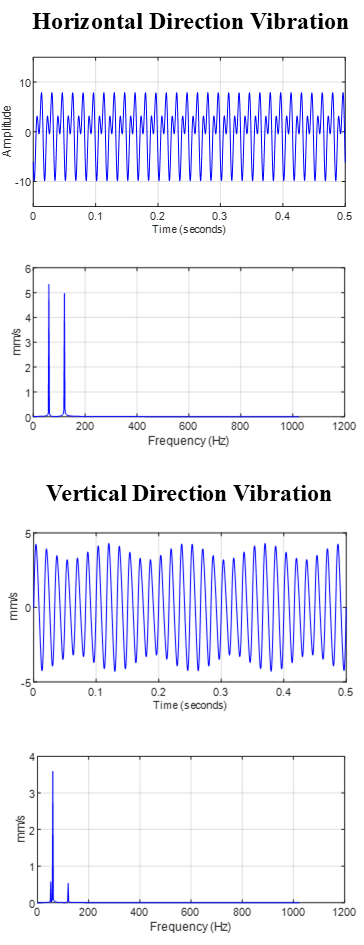} 
    \end{minipage}
\end{tcolorbox}

\captionof{figure}{Example 3.}
\label{fig:enter-label}

\begin{table}[H]
\centering
\caption{Diagnostic Results of Three Large Language Models}
\begin{tabular}{c l}
\hline
\textbf{Model} & \textbf{Answer} \\ \hline
Standard Answer & B. Misalignment \\ 
ChatGLM        & A. Unbalance \\ 
GPT-4          &    A. Unbalance \\
MaintAGT       & B. Misalignment \\ \hline 
\end{tabular}
\end{table}

\begin{tcolorbox}[colback=white, colframe=black, title=Evaluation results of multimodal large models., breakable,width=14cm]

\textbf{Answer(GPT4.0):}\\
The vibration patterns in the provided image show prominent frequency peaks around multiples of the rotational frequency (60 Hz). Such patterns are characteristic of certain types of equipment faults:
\\
\textbf{Unbalance} often manifests as a dominant vibration peak at the rotational frequency (60 Hz in this case).
\\
\textbf{Misalignment} can show as peaks at both the rotational frequency and its harmonics, though it may include additional vibration in multiple directions.
\\
\textbf{Looseness} typically presents with higher harmonics and a broader frequency response, often seen as a "noisy" or "spread-out" spectrum.
\\
\textbf{Blade Pass} would appear with peaks at a frequency related to the number of blades times the rotational frequency, which isn’t evident here.
\\
In this case, the vibration spectrum displays distinct peaks at the rotational frequency (60 Hz) and possibly its harmonics, which is most consistent with A. Unbalance as the likely fault type.\\

\textbf{Answer(ChatGLM4):} \\
The answer is A. Unbalance\\

\textbf{Answer(MaintAGT):}\\
B. Misalignment
\end{tcolorbox}
\captionof{figure}{The output of this model compared with the other two models.}
\label{fig:enter-label}
\medskip

Table 9 and Figure 10 present the output results of three large language models. The standard answer to the question is: "A: Unbalance." ChatGLM diagnosed the input signal as: "D: Resonance." This model ruled out unbalance and misalignment through simple feature analysis; however, due to its inability to conduct a more in-depth feature differentiation for modulation and resonance, it selected resonance as the most likely fault type. GPT-4 provided a diagnosis of: "B: Misalignment." GPT-4 primarily based its judgment on the analysis of spectral features, identifying the frequency components in the signal. However, it failed to consider the amplitude of individual frequencies during its analysis, leading to an inaccurate diagnosis. In contrast, the proposed MaintAGT model is capable of combining mathematical feature descriptions generated from the signal-to-text module with physical feature descriptions, and through chain-of-thought reasoning, it accurately diagnosed the signal as belonging to an unbalance condition of the equipment, providing the correct answer: "A: Unbalance." This result demonstrates the outstanding performance of MaintAGT in mechanical fault diagnosis, particularly in its ability to integrate multimodal information and physical understanding, surpassing other models.

\section{Conclusion}
\hspace*{2em}To address the deficiencies of large language models in specific domain knowledge and applications—particularly in the field of intelligent operations and maintenance—this study proposes MaintAGT, a professional large model for intelligent operations and maintenance. MaintAGT adopts the chain-of-thought paradigm, integrating both text and signal modalities to achieve comprehensive perception and intelligent diagnosis of mechanical equipment status. By incorporating an advanced signal-to-text conversion module and a text-based large model fine-tuned with specialized domain knowledge, we developed MaintAGT to enhance diagnostic accuracy and efficiency. Experimental results demonstrate that MaintAGT performs excellently across eight core tasks—including equipment condition monitoring, fault diagnosis, and maintenance recommendations—with accuracy and professionalism reaching the level of a human vibration analyst at ISO Level III.


\hspace*{2em}1) Superior Performance of the Signal-to-Text Model:In the single harmonic signal test, the model successfully captured the signal's frequency, amplitude, period, and initial phase, generating accurate textual descriptions.In the multi-harmonic signal test, it accurately identified the fundamental frequency and its second and third harmonics, providing detailed descriptions of their respective amplitudes and generating precise textual descriptions.Although there is room for improvement in describing the signal's noise components and nonlinear characteristics, the model overall demonstrates high accuracy and practicality, providing a reliable data foundation for advanced diagnostics in intelligent maintenance systems.The integration of signal and text modalities is a groundbreaking approach.

\hspace*{2em}2) Outstanding Performance of the Text Model:In the ISO Level III test questions covering eight key knowledge areas—condition monitoring, vibration principles, data acquisition, signal processing, fault analysis, corrective actions, equipment detection and diagnosis, and equipment knowledge—MaintAGT achieved an overall accuracy rate of 70\%, reaching the level of a human ISO Level III vibration analyst.
MaintAGT outperformed industry-leading large models like GPT, ChatGLM, and Qwen in all knowledge areas, exhibiting comprehensive and stable performance without significant weaknesses in any specific domain.

\hspace*{2em}3) Significant Advantages of the Multimodal Large Model:In the test of Multimodal Example 3, MaintAGT accurately diagnosed the equipment's fault type as "B. Misalignment," consistent with the standard answer.
MaintAGT is capable of integrating mathematical feature descriptions and physical characteristics of signals, utilizing chain-of-thought reasoning to accurately determine the equipment's fault type, surpassing the performance of models like GPT-4 and ChatGLM.
This result demonstrates MaintAGT's outstanding performance in mechanical fault diagnosis, particularly in multimodal information integration and physical understanding, proving its broad application prospects in the field of intelligent maintenance.

\hspace*{2em}Despite the excellent performance of the multimodal large model in many aspects, there are still some limitations. Firstly, the signal-to-text module requires optimization when processing complex signals, particularly in the description of noise and nonlinear components.Additionally, the model needs further refinement in the depth and breadth of specific professional knowledge.

\hspace*{2em}Future research will focus on enhancing the model's capabilities in specific professional knowledge to ensure its comprehensive reliability and precision in various application scenarios. We will further optimize the architecture and training strategies of MaintAGT, exploring new research directions such as task fine-tuning and human feedback mechanisms to improve the model's accuracy and practicality.

\clearpage
\appendix
\section{GLM-4's fine-tuning tips}
\label{app1}
\begin{tcolorbox}[colback=white, colframe=black, title=Fine-tuning Tip, breakable,width=14cm]
   
\textbf{Instruction:}  You are currently an excellent vibration analysis model, please answer the following questions:

\medskip
\textbf{Input:} What is meant by the term 'amplitude' in the context of vibration analysis? 

\medskip
\textbf{Output:} In vibration analysis, 'amplitude' refers to the magnitude or extent of the vibration waveform, typically measured as the maximum displacement from the equilibrium position.
\end{tcolorbox}

\section{Different types of training data}
\label{app2}
\begin{tcolorbox}[colback=white, colframe=black, title=Task Type and Example, breakable,width=14cm]
   
\centerline{\textbf{Condition Monitoring}}
\medskip
\textbf{Input:} Which one of the following combinations is detectable with ultrasound?  Select one correct answer from the list	A. Corona, arcing and tracking	B. Corona, 50Hz humming and tracking	C. Blasting, tracking and arcing	D. Looping, arcing and tracking 

\medskip
\textbf{Output:} A. Corona, arcing and tracking

\medskip
\centerline{\textbf{ Vibration Mechanism}}  
\medskip
\textbf{Input:} What is the difference between a beat signal and an amplitude modulated signal? Select one correct answer from the list	A. A beat frequency is generated when a big amplitude frequency mixes with a lower amplitude equal frequency, whereas amodulated signal is the result of an inner race defect on a bearing.	B. A beat frequency is the result of two closely spaced frequencies generated by the machine, whereas a modulated frequency is the result of a single frequency whose amplitude is changing periodically.	C. A beat freguency is when a blade pass freguency starts beating because of an unbalanced shaft, whereas a modulated frequency is when the blade pass frequency is affected by cavitation.	D. A beat frequency can occur when a flexible coupling gets old and starts to beat in amplitude, whereas amplitude modulation is the result of misalignment on the coupling.

\medskip
\textbf{Output:} B. A beat frequency is the result of two closely spaced frequencies generated by the machine, whereas a modulated frequency is the result of a single frequency whose amplitude is changing periodically.

\medskip    
\centerline{\textbf{Data Acquisition}}  
\medskip

\textbf{Input:} You performed a bump test on a beam and the waveform (a portion of impact signal) was obtained. You noticed that the waveform length is too short. Which of the following options would increase the waveform length? Select one correct answer from the list	A. Increase the sample rate, keeping the number of samples constant	B. Increase the number of samples, keeping the sample rate constant	C. Decrease the number of samples, keeping the sample rate constant	D. Increase the sample rate and decrease the number of samples

\medskip
\textbf{Output:} B. Increase the number of samples, keeping the sample rate constant

\medskip
\centerline{\textbf{Fault Analysis}} 
\medskip
\textbf{Input:} Which of the following failure modes can be detected using motor current signature analysis? Select one correct answer from the list	A. Broken rotor bars, shorted windings and earth leakage	B.  Stator eccentricity, bad connection and broken end rings	C. Broken rotor bars, rotor eccentricity and damaged end rings	D.  Shorted windings, stator and rotor eccentricity.

\medskip
\textbf{Output:} C. Broken rotor bars, rotor eccentricity and damaged end rings

\medskip
\centerline{\textbf{Corrective Action}} 
\medskip
\textbf{Input:} What is bar sag? Select one correct answer from the list	A. The brackets used by dial indicators and laser alignment systems are made of metal and will change length and even bends lightly if the machine being aligned is hot.	B. The rods used to support the dials will bend under the force of gravity. That bend adversely affects the accuracy of the readings	C. Bar sag results when bent shims are placed under the feet of the motor instead of new, straight shims. This condition is alsoknown as "squishy feet".	D. Bar sag is a term that used to be associated with older dial indicator alignment systems. Newer systems are better designedand no longer suffer from this problem.

\medskip
\textbf{Output:} B. The rods used to support the dials will bend under the force of gravity. That bend adversely affects the accuracy of the readings 

\medskip
\centerline{\textbf{Equipment Inspection and Diagnosis}}
\medskip
\textbf{Input:} Which of the following statements about proximity probe vibration measurement in journal bearings is correct? Select one correct answer from the list	A. A.C signal indicates the movement of the shaft, D.C signal indicates the distance between the shaft and the probe tip	B. D.C signal indicates the movement of the shaft, A.C signal indicates the position of the shaft	C. Proximity probes measure the bearing temperature in journal bearings 	D. Proximity probes can only measure the A.C signal

\medskip
\textbf{Output:} A. A.C signal indicates the movement of the shaft, D.C signal indicates the distance between the shaft and the probe tip 

\medskip
\centerline{\textbf{Equipment Knowledge}}
\medskip
\textbf{Input:} If after performing a number of modal test measurements we found that the coherence data was very poor at certain frequencies, what should you do? Select one correct answer from the list	A. Try impacting the structure at a different point (if the accelerometer is being moved around the structure). You may have chosen a nodal point.	B. Use more averages. lf you are already using four averages, try using between 6 and 10 averages	C. Unless you plan to perform structural dyamic modifications with the modal software you can ignore the coherence test data	D. You should repeat the test with the sensor oriented correctly.

\medskip
\textbf{Output:} A. Try impacting the structure at a different point (if the accelerometer is being moved around the structure). You may have chosen a nodal point. 
\end{tcolorbox}

\medskip
\medskip

\section{Evaluation Results of Large Text Models}
\begin{tcolorbox}[colback=white, colframe=black, title=Input: What is dynamic range?, breakable,width=14cm]
\textbf{Answer(MaintAGT):} Dynamic range is a measure of the ability to 'see' small signals in the presence of large ones. It is a ratio of the biggest to the smallest signal, and is typically represented in dB. \\
Dynamic range = 20 times log (Biggest signal/Smallest signal). A "wide" dynamic range means we can resolve very small signals in the presence of very large signals. The dynamic range is related to the number of bits in the A/D converter. To calculate it, raise 2 to the power of the number of bits in the A/D. If it has 12 bits, it would be $2^{12}$; this becomes the "biggest signal" in the formula above and the smallest signal is "1".\\

\textbf{Has concept, explanation, formula, and detailed examples}\\
\rule{\linewidth}{1pt}

\medskip

\textbf{Answer(GPT4.0):} Dynamic range is a measure of the difference between the smallest and largest values that a system can handle. In different contexts, it can refer to different things: \\
1. \textbf{Audio:} In audio engineering, dynamic range refers to the difference between the quietest and loudest sounds that can be accurately reproduced by a sound system or captured by a recording device. It is typically measured in decibels (dB). A higher dynamic range allows for more detailed and nuanced audio reproduction.\\
2. \textbf{Photography and Imaging:} In photography, dynamic range refers to the range of light intensities from the darkest shadows to the brightest highlights that a camera or sensor can capture. High dynamic range (HDR) photography techniques are used to capture scenes with a wide range of lighting conditions by combining multiple exposures.\\
3. \textbf{Electronics:} In electronics, dynamic range refers to the ratio between the smallest and largest possible signal levels in a system. This can apply to anything from digital-to-analog converters (DACs) to radio receivers. It indicates the range over which a device can accurately process signals.\\
4. \textbf{Physics and Engineering:} In general scientific and engineering contexts, dynamic range refers to the range over which a particular measurement system can provide reliable and accurate readings.\\
The concept of dynamic range is crucial in any field that involves capturing, processing, or reproducing signals, as it directly impacts the quality and fidelity of the information being handled.\\
5. \textbf{Physics and Engineering:} In general scientific and engineering contexts, dynamic range refers to the range over which a particular measurement system can provide reliable and accurate readings.\\
The concept of dynamic range is crucial in any field that involves capturing, processing, or reproducing signals, as it directly impacts the quality and fidelity of the information being handled.\\

\textbf{The answer is too broad and does not delve into the connotation of dynamic range for the field of intelligent operation and maintenance.}\\
\rule{\linewidth}{1pt}

\textbf{Answer(ChatGLM4):} Dynamic range refers to the difference between the smallest and largest possible values of a change in quantity. It is a measure of the range of values that can be represented, detected, or controlled. The concept of dynamic range is used in various fields, including physics, photography, audio recording, and more.\\
In photography, dynamic range refers to the range of light intensities that a camera sensor can capture from the darkest shadows to the brightest highlights, without losing detail. A camera with a high dynamic range can capture a wide range of brightness levels in a scene, resulting in an image with more detail in both the shadows and highlights.\\
In audio recording, dynamic range refers to the range of volume levels from the quietest to the loudest sound that can be accurately recorded and reproduced. A high dynamic range in audio allows for more accurate representation of the original sound, preserving subtle details and avoiding distortion.\\
In physics and other scientific fields, dynamic range may refer to the range of values that a measuring instrument can detect or the range of values that a system can handle without distortion or saturation.\\

\textbf{The definition is imprecise; ‘dynamic range’ refers to the ratio of the maximum signal to the minimum signal.} \\
\rule{\linewidth}{1pt}

\textbf{Answer(Qwen2.5):} Dynamic range refers to the ratio between the maximum and minimum values of a measurable quantity, such as sound or light. It describes the range of values over which a system can effectively capture or reproduce the variation in that quantity without distortion or loss of detail.\\
In photography and digital imaging, dynamic range refers to the range of brightness levels from the darkest shadows to the brightest highlights that a camera or display can capture or reproduce. A high dynamic range (HDR) means the device can capture both very bright and very dark areas with great detail, while a low dynamic range might result in loss of detail in the highlights or shadows.\\
In audio, dynamic range represents the difference between the quietest and loudest sounds that can be recorded or played back without distortion. A wide dynamic range allows for subtle details to be heard alongside powerful peaks.\\
Overall, a higher dynamic range provides a greater level of realism, detail, and contrast in both visual and audio media.\\

\textbf{The example provided is relatively simple and does not offer a comprehensive summary, as the specific content of the concept is not known without the given formula.}

\end{tcolorbox}


\begin{thebibliography}{99}
\bibitem{equipment_1}
Zang, Jiyuan and Liu, Yufei and Wang, Baicun and others. 
Technology forecasting and roadmapping of intelligent manufacturing by 2035. 
\textit{Journal of Mechanical Engineering}, 2022, 58(4), 285-308.

\bibitem{equipment_2}
Wang, Junliang and Gao, Pengjie and Zhang, Jie and Wang, Lihui and others. 
A review of manufacturing big data: Connotation, methodology, application and trends. 
\textit{Journal of Mechanical Engineering}, 2023, 59(12), 1-16.

\bibitem{damage_3}
Tauheed, Mian and Choudhary, Anurag and Shahab, Fatima. 
Multi-sensor fault diagnosis for misalignment and unbalance detection using machine learning. 
\textit{IEEE Transactions on Industry Applications}, 2023, 59(5), 5749-5759.

\bibitem{damage_4}
Lei, Yaguo and Yang, Bin and Jiang, Xinwei and others. 
Applications of machine learning to machine fault diagnosis: A review and roadmap. 
\textit{Mechanical Systems and Signal Processing}, 2020, 138, 106587.

\bibitem{damage_5}
Chen, Jiaxian and Huang, Ruyi and Chen, Zhuyun and others. 
Transfer learning algorithms for bearing remaining useful life prediction: A comprehensive review from an industrial application perspective. 
\textit{Mechanical Systems and Signal Processing}, 2023, 193, 110239.

\bibitem{damage_6}
Huang, Baoyu and Zhang, Yongxiang. 
Rolling element bearing fault diagnosis using a three-step scheme. 
\textit{Journal of Mechanical Engineering}, 2024, 60(14).

\bibitem{safety_7}
Ren, Shan and Wang, Jin and Zhao, Xin and others. 
“Doubly-fed” manufacturing service of intelligent design and preventive maintenance for complex products. 
\textit{Journal of Mechanical Engineering}, 2024, 60(6), 127-136.

\bibitem{safety_8}
Zheng, Zhe and Wang, Fei and Gong, Guofang and others. 
Intelligent technologies for construction machinery using data-driven methods. 
\textit{Automation in Construction}, 2023, 147, 104711.



\bibitem{big_data_10}
Lei, Yaguo and Jia, Feng and Kong, Detong and others. 
Opportunities and challenges of machinery intelligent fault diagnosis in big data era. 
\textit{Journal of Mechanical Engineering}, 2018, 54(5), 94-104.

\bibitem{Zhang_11}
Xu, Haifeng and Wang, Xu and Huang, Jinfeng and others. 
Semi-supervised multi-sensor information fusion tailored graph embedded low-rank tensor learning machine under extremely low labeled rate. 
\textit{Information Fusion}, 2024, 105, 102222.

\bibitem{domain_adversarial_training_12}
Li, Yibin and Song, Yan and Jia, Lei and others. 
Intelligent fault diagnosis by fusing domain adversarial training and maximum mean discrepancy via ensemble learning. 
\textit{IEEE Transactions on Industrial Informatics}, 2021, 17(4), 2833-2841.

\bibitem{deep_adversarial_domain_13}
Liu, Zhaohua and Lu, Biliang and Wei, Hualiang and others. 
Deep adversarial domain adaptation model for bearing fault diagnosis. 
\textit{IEEE Transactions on Systems, Man, and Cybernetics: Systems}, 2021, 51(7), 4217-4226.



\bibitem{llms1}
Yang, Jingfeng and Jin, Hongye and Tang, Ruixiang and others. 
Harnessing the power of LLMs in practice: A survey on ChatGPT and beyond. 
\textit{ACM Transactions on Knowledge Discovery from Data}, 2024, 18(6), 1-32.

\bibitem{llms2}
Zhao, Wayne Xin and Zhou, Kun and Li, Junyi and others. 
A survey of large language models. 
\textit{arXiv preprint arXiv:2303.18223}, 2023.

\bibitem{chatgpt3}
Brown, Tom and Mann, Benjamin and Ryder, Nick and others. 
Language models are few-shot learners. 
\textit{Advances in Neural Information Processing Systems}, 2020, 33, 1877-1901.

\bibitem{chatgpt4}
Radford, Alec and Wu, Jeffrey and Child, Rewon and others. 
Language models are unsupervised multitask learners. 
\textit{OpenAI blog}, 2019, 1(8), 9.

\bibitem{chatgpt5}
Radford, Alec and Narasimhan, Karthik and Salimans, Tim and others. 
Improving language understanding by generative pre-training. 
Unpublished manuscript, 2018.

\bibitem{gpt-4_6}
Bubeck, Sébastien and Chandrasekaran, Varun and Eldan, Ronen and others. 
Sparks of artificial general intelligence: Early experiments with GPT-4. 
\textit{arXiv preprint arXiv:2303.12712}, 2023.

\bibitem{llama7}
Touvron, Hugo and Martin, Louis and Stone, Kevin and others. 
LLaMA 2: Open foundation and fine-tuned chat models. 
\textit{arXiv preprint arXiv:2307.09288}, 2023.

\bibitem{llama8}
Meta AI. 
Introducing Meta LLaMA 3: The most capable openly available LLM to date. 
\textit{Meta AI}, 2024.

\bibitem{llama9}
Touvron, Hugo and Lavril, Thibaut and Izacard, Gautier and others. 
Llama: Open and efficient foundation language models. 
\textit{arXiv preprint arXiv:2302.13971}, 2023.

\bibitem{bloomberg_gpt}
Wu, Shijie and Irsoy, Ozan and Lu, Steven and others. 
BloombergGPT: A Large Language Model for Finance. 
\textit{arXiv preprint arXiv:2303.17564}, 2023.

\bibitem{law}
Lai, Jinqi and Gan, Wensheng and Wu, Jiayang and others. 
Large Language Models in Law: A Survey. 
\textit{arXiv preprint arXiv:2312.03718}, 2023.

\bibitem{biogpt}
Luo, Renqian and Sun, Liai and Xia, Yingce and others. 
BioGPT: Generative Pre-trained Transformer for Biomedical Text Generation and Mining. 
\textit{Briefings in Bioinformatics}, 23(6): bbac409, 2022.

\bibitem{sun}
Sun, Y., Zhang, Q., Bao, J., and others. 
Empowering digital twins with large language models for global temporal feature learning. 
\textit{Journal of Manufacturing Systems}, 74: 83-99, 2024.


\bibitem{xia}
Xia, Liqiao and Li, Chengxi and Zhang, Canbin and others. 
Leveraging error-assisted fine-tuning large language models for manufacturing excellence. 
\textit{Robotics and Computer-Integrated Manufacturing}, 2024, 88, 102728.

\bibitem{zheng}
Zheng, Shuwen and Pan, Kai and Liu, Jie and others. 
Empirical study on fine-tuning pre-trained large language models for fault diagnosis of complex systems. 
\textit{Reliability Engineering and System Safety}, 2024, 252, 110382.

\bibitem{lowenmark}
Lowenmark, Karl and Taal, Cees and Schnabel, Stephan and others. 
Technical language supervision for intelligent fault diagnosis in process industry. 
\textit{arXiv preprint arXiv:2112.07356}, 2021.

\bibitem{signal_representation_learning}
Wang, Huan and Liu, Zhiliang and Ge, Yipei and others. 
Self-supervised signal representation learning for machinery fault diagnosis under limited annotation data. 
\textit{Knowledge-Based Systems}, 2022, 239, 107978.

\bibitem{CoT_1}
Wei, Jason and Wang, Xuezhi and Schuurmans, Dale and others. 
Chain-of-thought prompting elicits reasoning in large language models. 
\textit{Advances in Neural Information Processing Systems}, 2022, 35, 24824-24837.

\bibitem{CoT_2}
Li, Jiachun and Cao, Pengfei and Chen, Yubo and others. 
Towards faithful chain-of-thought: Large language models are bridging reasoners. 
\textit{arXiv preprint arXiv:2405.18915}, 2024.

\bibitem{CoT_3}
Kojima, Takeshi, Gu, Shixiang, Reid, Mark, and others. 
Large Language Models are Zero-Shot Reasoners. 
\textit{Advances in Neural Information Processing Systems}, 35: 22199-22213, 2022.

\bibitem{CoT_4}
Zhang, Zhuosheng, Zhang, Aston, Li, Mu, and others. 
Automatic Chain of Thought Prompting in Large Language Models. 
\textit{arXiv preprint arXiv:2210.03493}, 2022.

\bibitem{imbalance_1}
Xu, Haifeng, Wang, Xu, Huang, Jinfeng, Zhang, Feibin, and Chu, Fulei. 
Semi-supervised multi-sensor information fusion tailored graph embedded low-rank tensor learning machine under extremely low labeled rate. 
\textit{Information Fusion}, 105: 102222, 2024.

\bibitem{imbalance_2}
Huang, Jinfeng, Zhang, Feibin, Safaei, Babak, and others. 
The Flexible Tensor Singular Value Decomposition and Its Applications in Multisensor Signal Fusion Processing. 
\textit{Mechanical Systems and Signal Processing}, 220: 111662, 2024.

\bibitem{efficiency_1}
Korkos, Panagiotis, Linjama, Matti, Kleemola, Jaakko, and Lehtovaara, Arto. 
Data annotation and feature extraction in fault detection in a wind turbine hydraulic pitch system. 
\textit{Renewable Energy}, 185: 692-703, 2022.

\bibitem{efficiency_2}
Wang, Huan, Liu, Zhiliang, Ge, Yipei, and Peng, Dandan. 
Self-supervised signal representation learning for machinery fault diagnosis under limited annotation data. 
\textit{Knowledge-Based Systems}, 239: 107978, 2022.

\bibitem{glm4}
GLM, Team, Zeng, Aohan, Xu, Bin, and others. 
ChatGLM: A Family of Large Language Models from GLM-130B to GLM-4 All Tools. 
\textit{arXiv preprint arXiv:2406.12793}, 2024.


\end{thebibliography}
\end{document}